\documentclass[11pt,english,twoside]{article}

\usepackage[T1]{fontenc}
\usepackage[latin1]{inputenc}
\usepackage[english]{babel}
\usepackage{lmodern}
\usepackage{a4wide}
\usepackage{amssymb, amsmath, amsthm}
\usepackage{slashed}
\usepackage{float}
\usepackage{graphicx}
\usepackage[dvips]{epsfig}
\usepackage{psfrag}
\usepackage{lscape}
\usepackage[all]{xy}
\usepackage{hyperref}
\usepackage{enumerate}
\usepackage{dsfont}

\newcommand{\beq}{\begin{equation}}
\newcommand{\eeq}{\end{equation}}
\def\bea#1\eea{\begin{align}#1\end{align}} 
\newcommand{\nn}{\nonumber}
\newcommand{\w}{\wedge}
\newcommand{\del}{\partial}
\newcommand{\id}{\mathds{1}}

\def\d {{\rm d}}
\def\bbb{{\cal B}}
\def\fff{{\cal F}}
\def\lll{{\cal L}}
\def\mmm{{\cal M}}
\def\eee{{\cal E}}
\def\hhh{{\cal H}}
\def\ddd{{\cal D}}
\def\teee{\tilde{{\cal E}}}
\def\tL{\tilde{{\cal L}}}
\def\te{\tilde{e}}
\def\tf{\tilde{f}}
\def\tg{\tilde{g}}

\def\tp {\tilde{\phi}}

\def\tb {\tilde{\beta}}
\def\tQ {\tilde{Q}}
\def\he{\hat{e}}
\def\hg{\hat{g}}
\def\hB {\hat{B}}
\def\hH {\hat{H}}
\def\hp {\hat{\phi}}
\def\heee{\hat{{\cal E}}}
\def\hL{\hat{{\cal L}}}
\def\iG {(G^{-1})}
\def\G {\Gamma}

\def\vs {\vphantom{\smash[t]{\Big(}}}
\def\R {\mathcal{R}}
\def\tR{\widetilde{\mathcal{R}}}
\def\hR{\widehat{\mathcal{R}}}

\makeatletter
\@addtoreset{equation}{section}
\makeatother

\begin{document}

\begin{titlepage}

\rightline{\small LMU-ASC 19/11}
\rightline{\small MPP-2011-64}

\vskip 2.8cm 

{\fontsize{18.2}{21}\selectfont 
  \flushleft{\noindent\textbf{A ten-dimensional action for non-geometric fluxes}} }

\vskip 0.2cm
\noindent\rule[1ex]{\textwidth}{1pt}
\vskip 1.3cm

\noindent\textbf{David Andriot$^{a}$, Magdalena Larfors$^{a}$, Dieter L\"ust$^{a,b}$, Peter Patalong$^{a,b}$}

\vskip 0.6cm
\begin{enumerate}[$^a$]
\item \textit{Arnold-Sommerfeld-Center for Theoretical Physics\\Department f\"ur Physik, Ludwig-Maximilians-Universit\"at M\"unchen\\Theresienstra\ss e 37, 80333 M\"unchen, Germany}
\vskip 0.2cm
\item \textit{Max-Planck-Institut f\"ur Physik\\F\"ohringer Ring 6, 80805 M\"unchen, Germany}
\end{enumerate}
%
\noindent {\small{\texttt{david.andriot@physik.uni-muenchen.de, magdalena.larfors@physik.uni-muenchen.de,\\dieter.luest@lmu.de, peter.patalong@physik.uni-muenchen.de}}}

\vskip 2.5cm

\begin{center}
{\bf Abstract}
\end{center}

\noindent The NSNS Lagrangian of ten-dimensional supergravity is rewritten via a change of field variables inspired by Generalized Complex Geometry. We obtain a new metric and dilaton, together  with an antisymmetric bivector field which leads to a ten-dimensional version of the non-geometric $Q$-flux. Given the involved global aspects of non-geometric situations, we prescribe to use this new Lagrangian, whose associated action is well-defined in some examples investigated here. This allows us to perform a standard dimensional reduction and to recover the usual contribution of the $Q$-flux to the four-dimensional scalar potential. An extension of this work to include the $R$-flux is discussed. The paper also contains a brief review on non-geometry.

\vfill

\end{titlepage}

\tableofcontents
\newpage


\section{Introduction}

Among the rich structures discovered in string theories, some of the most intriguing ones are ``stringy'' symmetries, which have no correspondence in point particle theories. T-duality is an example of such, for some backgrounds. These additional symmetries can serve various purposes, like constructing an orbifold, or patching the target space fields of a string configuration,\footnote{We often prefer the terms ``configuration'', ``set-up'' or ``situation'', to a ten-dimensional supergravity non-geometric ``solution'' or ``background''. Indeed, most of the time we only consider a set of Neveu-Schwarz (NSNS) fields, that must be completed by other ingredients to solve the equations of motion. Usually, NSNS fields alone are not enough to get a compact solution.} instead of using the standard geometric transition functions (diffeomorphisms and gauge transformations). Doing so creates unusual situations with respect to point-like geometries. These ideas \cite{hmw02, dh02} have lead to the development of non-geometry, which has been explored through many different angles in the literature. We provide a brief review in section \ref{sec:bckgd}.

Non-geometry is of particular interest for constructing phenomenologically interesting models in four-dimensional supergravities. In this setting, non-geometry appears through so-called non-geometric $Q$- and $R$-fluxes \cite{stw05}. These fluxes generate terms in the potential, that lead to desired phenomenological features. For instance, they sometimes help to fix all moduli \cite{stw06, mpt07, p07}. They also contribute positively to the four-dimensional cosmological constant (a rather non-standard behaviour in type II supergravities), and this helps to obtain de Sitter solutions \cite{cgm09a, cgm09b}. Furthermore, an interesting relation between non-geometrical fluxes and non-commutative and non-associative
closed string geometry was discovered recently \cite{Blumenhagen:2010hj, Lust:2010iy, Blumenhagen:2011ph}, see also \cite{gs06} and references therein.

Four-dimensional supergravities with non-geometric potential terms have been identified as gauged supergravities, and the $Q$- and $R$-fluxes appear as some of the structure constants of the corresponding gauge algebras  \cite{dh02, stw05, dh05}. The other structure constants of these algebras usually have their origin in ten-dimensional supergravity fields, like the Neveu--Schwarz (NSNS) flux or the metric (curvature);  accordingly, compactifications from ten to four dimensions can lead to four-dimensional supergravities,
where the terms in the potential can be traced back to some ten-dimensional supergravity origin. This is unfortunately not the case for the non-geometric fluxes.

In ten-dimensional supergravity, non-geometry appears in a different fashion. When going around a non-contractible loop in target space, one can patch the supergravity fields using a stringy symmetry. Even if the fields are then globally well-defined from the string point of view, they do not appear single-valued because they are not glued with the standard geometric transition functions. One then deviates from the usual geometric set-up, and so the term ``non-geometry'' is used \cite{hmw02, fww04}. This is the simplest non-geometric situation, which is thought to correspond to the presence of a $Q$-flux. Note that this point of view on non-geometry does not involve any other degree of freedom than the supergravity fields. In particular there is a priori no appearance of some new field which could correspond to the non-geometric fluxes.

Put differently, no straight connection (via a dimensional reduction for instance) has been established between the ten- and four-dimensional supergravity points of view on non-geometry. This forbids the lifting of some four-dimensional gauged supergravities, and their solutions, to ten-dimensional ones; more generally, this prevents us from getting an understanding of the non-geometric fluxes, and the associated phenomenological models, in terms of ten-dimensional (supergravity) target space fields. In this paper, we make some progress in this direction, and clarify the relation between ten- and four-dimensional non-geometry, by finding a way to make the non-geometric fluxes appear in ten-dimensional supergravity. Before we describe this result in detail, let us briefly motivate it.\\

In the ten-dimensional point of view just described, one assumes a local geometry,\footnote{This non-geometric situation, which remains locally geometric, should give rise to a $Q$-flux. One can also face a set-up where the local notion of geometry is lost \cite{dh05, stw06}. An $R$-flux should then also be present. However, we will not consider $R$-fluxes in this paper.} and the non-geometry arises as a global issue: it is only when gluing the fields across a few patches, that one must deviate from standard geometric transition functions. Consequently, it is  only globally that fields  appear (naively) ill-defined. These global issues could be a first reason to believe that ten-dimensional supergravity is not suited to describe non-geometry.  Indeed, its action assumes the point-like $\alpha'\rightarrow 0$ limit of string theory and the standard integration of fields, which do not accommodate the stringy aspects of the geometry. In other words, the ten-dimensional supergravity effective description may not be enough to capture what looks like stringy effects. Possibly, a proper description would instead be provided by some string construction. Asymmetric orbifolds give examples of such stringy descriptions, that lift some (non-geometric) gauged supergravities \cite{dh02, dh05}. More concretely, these questions on global issues and integration seem to forbid a standard compactification procedure. Therefore, it looks difficult to establish a direct relation with the  four-dimensional non-geometric fluxes.
 
However, we can make some remarks on the non-geometric fluxes and their possible higher dimensional origin. In four dimensions, the $Q$- and $R$-fluxes were at first motivated by asking for T-duality covariance of the superpotential and the gauge algebras in the NSNS sector. A way to understand this is that in ten dimensions, it was observed \cite{kstt02} that some non-geometric situations could be reached by applying T-dualities on geometric situations; the same should then occur in four dimensions. A well-known example of a non-geometric configuration obtained in this way is given by a two-torus $T^2$ fibered over a base circle, which we come back to in detail in appendix \ref{ap:toyex}. In this toroidal example, and many others, T-duality is also the stringy symmetry used to make fields single-valued in ten dimensions. Note that T-duality (at least in its standard form) is a symmetry only for string backgrounds with some isometries. In the case of the toroidal example the isometries are along the $T^2$ directions.

In this paper, we will be inspired by the toroidal example and the role of T-duality, even if our results also apply more generically. Therefore, we will focus only on the NSNS sector of ten-dimensional supergravity.\footnote{T-duality does not mix sectors, so if the $Q$- and $R$- fluxes arise from the NSNS sector and not the Ramond-Ramond (RR) sector, studying the former is enough. In addition, the incorporation of RR-fluxes in type II non-geometric situations is always complicated by their possible D-brane or O-plane sources, that are difficult to fit with the non-geometric global aspects. For simplicity, we will not consider $N\!S5$-branes either. Let us note however some discussions of non-geometric solutions of heterotic string \cite{fww04, rs08, mms10, a11}.}
As explained above, the non-geometric fluxes should then arise by applying T-duality on the NSNS fields. However, in ten dimensions, T-dualising some geometric configuration like the toroidal example leads to the NSNS fields becoming ill-defined. No new degree of freedom appears. This is a first hint that the non-geometric fluxes do not correspond to any new degree of freedom, present in some more advanced theory, or in some involved string construction, but they are just given by the NSNS fields.

With this in mind, we turn our attention to the NSNS action. It can be rewritten in a T-duality covariant way \cite{ms92, hhz10}, up to a total derivative term which we will come back to (in particular, we come back to double field theory in section \ref{sec:DFT}). Such a rewriting is possible thanks to the introduction of an object called the generalized metric $\hhh$, which depends on the metric $\hg$ and the Kalb-Ramond $B$-field $\hB$. For a $d$-dimensional space-time, $\hhh$ is the $2d \times 2d$ matrix with $d \times d$-blocks\footnote{The unconventional notation $\hg$ and $\hB$ is chosen for later convenience, since we will soon consider fields in different bases. Our conventions for the metric and form fields are given in appendix \ref{ap:conv} and $\hhh$ is also defined in section \ref{sec:nongeomGCG}.}
\beq
\hhh=\begin{pmatrix} \hg-\hB \hg^{-1}\hB & \hB \hg^{-1} \\ -\hg^{-1} \hB & \hg^{-1}\end{pmatrix} \ .
\eeq
The T-duality covariant rewriting of the NSNS action is done in terms of $\hhh$ and the dilaton $\hp$. Such a rewriting implies that, if we perform a T-duality which brings the fields to a non-geometric situation (for instance, $\hg$ and $\hB$ in $\hhh$ are ill-defined), the resulting action would still be the NSNS action. So again, even if the ten-dimensional NSNS action could be argued to be ill-defined because of global properties, it looks like no other degree of freedom would appear, which could provide non-geometric fluxes.\\

These observations lead us to think that the non-geometric fluxes could appear in ten dimensions via a rewriting of the NSNS action. The NSNS degrees of freedom would be the only ones involved, but they would be mixed into new variables which would reveal non-geometric fluxes in the action. To find the good variables, we were inspired by the Generalized Complex Geometry (GCG) \cite{h02, g04} approach, where a particular object has been related \cite{gs06, mpt07, gs07, gmpw08} to non-geometry: the antisymmetric bivector $\beta^{mn}$. We come back to these ideas in detail in section \ref{sec:bckgd}. Note that this rewriting will give us new tools to discuss the global issues of the action, and the dimensional reduction.

The starting point is that the generalized metric $\hhh$ can, similarly to a usual metric, be expressed in terms of generalized vielbeine. 
A positive-definite metric, given by a $d\times d$ matrix $g$, can be written as $g=e^T \id_d\ e$ in terms  of the vielbein matrix $e$. However, the same metric could equally well be obtained from the vielbein
 $k e$, with $k \in O(d)$. The form of the vielbein is therefore not unique. Similarly, the generalized metric $\hhh$ can be expressed as
 $\hhh=\eee^T \id_{2d}\ \eee$. Let us introduce two possible generalized vielbeine
\beq
\heee=\begin{pmatrix} \he & 0 \\ -\he^{-T}\hB & \he^{-T} \end{pmatrix} \quad , \quad \teee=\begin{pmatrix} \te & \te \tb \\ 0 & \te^{-T} \end{pmatrix} \ .
\eeq
We will refer to these different forms of generalized vielbeine, and to their content, as vielbeine or fields in different ``bases'', in particular the hatted and tilded bases. We leave the word ``frame'' for the case where a T-duality is applied, as for instance in appendix \ref{ap:toyex}. $\heee$ is a natural choice for the generalized vielbein, since it corresponds to the metric  $\hg=\he^T \id_d\ \he$ and the $B$-field $\hB$ which appear in $\hhh$. However, a different generalized vielbein is a priori possible. In particular, a transformation like a T-duality or an $O(2d)$ action on a generalized vielbein sometimes results in one of the form of $\teee$, whose top right corner is given by an antisymmetric bivector $\tb^{mn}$. For various arguments reviewed in section \ref{sec:bckgd}, the appearance of such a $\beta$ block can be a sign of non-geometry. In addition, reproducing relations analogous to those of the gauge algebras of gauged supergravities, expressions for the non-geometric fluxes $Q$ and $R$ in terms of $\beta$ were proposed in \cite{gmpw08} (see \eqref{nongeoflux}), again showing a relation between $\beta$ and non-geometry.

These results inspired us the way to rewrite the NSNS action: in order to make the non-geometric fluxes appear, we should relate the standard NSNS fields to $\beta$, which seems to be the good
 variable for non-geometry. $\beta$ appears by considering the generalized vielbein in the tilded basis $\teee$, rather than the standard $\heee$. However, both give the same generalized metric:
\beq
\hhh=\begin{pmatrix} \hg-\hB \hg^{-1}\hB & \hB \hg^{-1} \\ -\hg^{-1} \hB & \hg^{-1}\end{pmatrix} = \heee^T \id_{2d}\ \heee = \teee^T \id_{2d}\ \teee 
= \begin{pmatrix} \tg & \tg\tb \\ -\tb \tg & \tg^{-1}-\tb \tg \tb \end{pmatrix} \ , \label{eq:intequality}
\eeq
where $\tg=\te^T \id_d\ \te$ is a new metric defined with respect to the $\te$ vielbein in the tilded basis.\footnote{\label{foot:sign}For metrics with Lorentzian signature, one should change $\id_d$ into the Minkowski metric, and the matrix $\id_{2d}$ accordingly. This does not change the result of \eqref{eq:intequality}.} From the relation \eqref{eq:intequality}, together with a definition of the dilaton $\tp$ that leaves the measure of the NSNS action invariant, we read off the change of variables from the  fields  $\hg$, $\hB$ and $\hp$  to the fields $\tg$, $\tb$, and $\tp$. It is then reasonable to assume that the NSNS Lagrangian can equally well be written in the new variables. 

In this paper, we focus for simplicity on the locally geometric situation, and do not consider any $R$-flux. A way to implement this is to make the assumption 
\beq
\tb^{km}\partial_m\cdot=0 \ , \label{eq:intassume}
\eeq
where the dot is a placeholder for any field. While we motivate this assumption in more detail in section \ref{sec:computation}, let us  for now simply mention that \eqref{eq:intassume} relies on the idea that the $\beta$-field should be along the isometry directions, as it is the case in the toroidal example. Given this assumption, the expressions proposed for the non-geometric fluxes \cite{gmpw08} with curved space indices become
\beq
\label{eq:nongeofluxsimple}
{Q_m}^{np}=\del_m \tb^{np} \ , \ R^{mnp}=0 \ .
\eeq
Using the assumption, we prove in section \ref{sec:computation} that
\beq
\hL= e^{-2\hp} \sqrt{|\hg|} \left(\hR + 4|\d \hp|^2 - \frac{1}{2} |\hH|^2 \right)= e^{-2\tp} \sqrt{|\tg|} \left(\tR + 4|\d \tp|^2 - \frac{1}{2} |Q|^2 \right) + \del (\dots) = \tL + \del (\dots) \ , \label{eq:intlag}
\eeq
where $\hL$ is the standard NSNS Lagrangian, and $\tL$ is the Lagrangian given by the tilded fields, with a $Q$-flux term. The metric used for the squares in $\hL$ is $\hg$, while it is $\tg$ in $\tL$ and the $H$-flux is given by the three-form $\hH=\d \hB$ (see appendix \ref{ap:conv} for further conventions). We show that the Lagrangians $\hL$ and $\tL$ with fields in the two different bases are equal up to a total derivative term. The assumption \eqref{eq:intassume} simplifies the derivation of \eqref{eq:intlag} considerably, and we leave the full computation for future work \cite{a11b}.\\

Given the rewriting \eqref{eq:intlag}, it is tempting to trade the NSNS action for the action associated to $\tL$. However, the integration of these Lagrangians makes us face the global issues mentioned in non-geometric situations. We discuss those at length in section \ref{sec:totder}, together with the fate of the total derivative term. It turns out that the change of variables helps us to propose a way out from these global issues. Indeed, $\tg$, $\tp$ and $Q$ can actually be well-defined in some non-geometric situations like the toroidal example, whereas $\hg$, $\hp$ and $\hH$ are ill-defined. We argue that this is an illustration of a more general situation, where one can find a preferred basis for the fields which leads to a single-valued Lagrangian. In addition, it should differ from the NSNS one by at most a total derivative term, as in \eqref{eq:intlag}. Then, we propose as a prescription that the associated action describes the low-energy physics properly. Following this idea, we use the well-defined action obtained by integrating $\tL$, and perform a dimensional reduction of it. This allows us to finally draw the link with the non-geometric fluxes appearing in the four-dimensional scalar potential.\\

The paper is organized as follows. Section \ref{sec:bckgd} contains a brief review on non-geometry, together with a focus on GCG. This serves as a background and motivation for the rewriting of the NSNS Lagrangian, which is performed in section \ref{sec:computation}. We also discuss there the assumption \eqref{eq:intassume} that we use. In section \ref{sec:disc}, we turn to the global aspects of the Lagrangians $\hL$ and $\tL$, the resulting actions and the total derivative term. We argue in favour of the prescription just mentioned and then focus on the action associated to $\tL$. This rewritten Lagrangian is then dimensionally reduced to four dimensions, and the result is shown to share essential properties with four-dimensional non-geometric models existing in the literature. The ten-dimensional equations of motion are also derived from the Lagrangian, and the Bianchi identity of the $Q$-flux is discussed. We end this section with a proposal on a double field theory Lagrangian for non-geometry. Our conclusions and outlooks for future work are presented in section \ref{sec:conclusion}. Three appendices complement the paper. Appendix \ref{ap:conv} gives the conventions used in the paper. Appendix \ref{ap:toyex} contains a detailed description of well-known geometric and non-geometric toroidal fibrations that are related by T-duality. This serves as an illustration for the relevance of different field bases in geometric and non-geometric settings. Appendix \ref{ap:comp} contains details of the computation performed in section \ref{sec:computation}.

\section{A brief review on non-geometry}\label{sec:bckgd}

In this section, we come back in more detail to the literature on non-geometry, in order to further motivate the result discussed in the introduction. We start with a brief review on non-geometry, where we do not intend to be exhaustive but focus on the aspects we need later on. After mentioning shortly the different aspects of non-geometry in various dimensions, we focus on the Generalized Complex Geometry perspective along the lines of \cite{gmpw08}. Given this background material, we return to the main computation and result of this paper in the next section.

\subsection{Non-geometry in various dimensions}\label{sec:rev}

The initial idea of non-geometry \cite{hmw02, dh02} is based on the fact that string theory has more symmetries than point particle theories do. These extra (stringy) symmetries could be used to build new string vacua and associated four-dimensional effective theories, which could be of particular interest for phenomenology. 

From a target space point of view, the extra symmetries play a role when constructing non-geometric string configurations.  In a given space, when going from one patch to another, fields of a string background are usually glued by diffeomorphisms and gauge transformations. However, it could be argued that a string background is still obtained if another string symmetry is used for this gluing, in particular a more stringy one like a duality. Since diffeomorphisms and gauge transformations are the only allowed transition functions in standard geometric settings involving smooth manifolds, one would talk of non-geometry for any other transformation.

In practice, the fields of a non-geometric configuration appear ill-defined when going around a loop in target space, but become single-valued only when allowing the additional string symmetry as a transition function. For such a monodromy to occur, one would need a non-contractible loop. The first non-geometric vacua with duality monodromies were obtained on a manifold with a singularity \cite{hmw02} or with a non-trivial one-cycle \cite{lnr03, fww04}. Concretely, one considers a fibration of a space $\fff$ over a base $\bbb$, so that while going around a non-contractible loop on $\bbb$, the fiber fields are glued from one patch to the other using the symmetries of string theory on $\fff$. The
 case of a torus fiber, $\fff=T^n$, with fields independent of the fiber coordinates (i.e. there are $n$ isometries), is of particular interest. String theory on such a fiber admits an important symmetry group, that has the T-duality group $O(n,n,\mathbb{Z})$ as a subgroup (for a review see \cite{gpr94, t10}).\footnote{The T-duality group is the continuous $O(n,n,\mathbb{R})$ at the supergravity level, but is broken to its discrete version in string theory. For simplicity in the following, we will only talk of $O(n,n)$.} This symmetry group can then be used to patch the fiber fields. In appendix \ref{ap:toyex}, we illustrate that situation with the well-known toroidal example \cite{kstt02, lnr03} of a $T^2$ fiber over a base circle, where the fiber components of the metric and the $B$-field are patched by a T-duality transformation. In this paper, we will focus on T-duality, but other string symmetries like S- and U-duality, or mirror symmetry,\footnote{Note that the  SYZ conjecture \cite{syz96} implies that for Calabi-Yaus that have a toroidal fiber, mirror symmetry is equivalent to T-duality.} could also be considered \cite{kv96, hc03, r06}.

There are several arguments for why non-geometric configurations like those just described should be consistent backgrounds of (at least classical) string theory \cite{h04} (see \cite{hm06, h06b, bct07, bt07, c11} for related discussions on the quantized string). The string conformal field theory (CFT) should remain the same under string symmetries, even if its representation as a sigma model in terms of target space fields could change (e.g. under T-duality). Furthermore, it was quickly noticed that some non-geometric backgrounds can be obtained from string geometric backgrounds by applying T-dualities \cite{kstt02, lnr03}, or equivalently \cite{syz96} mirror symmetry \cite{glmw02, fmt03, gs07}.\footnote{Even if some non-geometric backgrounds are obtained out of geometric ones using dualities, it is not a generic feature (see for instance \cite{st08, stw06}).} In this case, starting with a consistent geometric background would guarantee that quantum constraints like modular invariance are also satisfied for the dual non-geometric background \cite{h04}. It is then of interest to provide a string world-sheet theory on such backgrounds. This has been realized only in a few cases \cite{dh02, hmw02, fw05, dh05, hw06}, using the same initial idea of the stringy symmetries, to perform ``duality twists'' \cite{dh02}. The result is an asymmetric orbifold \cite{mw86, nsv87} or a deformation of such.\\

The ten-dimensional target space picture of non-geometry was thus initially that of naively ill-defined fields, which could be made globally defined only via non-standard stringy transitions functions. The four-dimensional picture appeared in a very different fashion (for a review see \cite{w07}). In a geometric configuration, the NSNS sector contributes to the four-dimensional superpotential via $H_{abc}$ (components of the $H$-flux in flat space indices), and $f^a_{\ bc}$ (structure constants obtained from derivatives of vielbeine as in \eqref{struct}, also known as ``geometric fluxes''; they correspond to the curvature contribution). We mentioned previously that some non-geometric configurations could arise by applying T-duality on geometric cases. With this idea in mind, a T-duality covariant expression generalizing the four-dimensional superpotential  was proposed in \cite{stw05}. This generalized superpotential has new contributions from ``non-geometric'' $Q$- and $R$-fluxes. The index structure of these fluxes can be inferred by considering T-duality rules that generalize the Buscher rules \cite{b87, b88} of the toroidal example, and are also inspired from RR fluxes (see \cite{stw05} and references therein). Concretely, one lifts or lowers the index
 in the T-duality direction
\beq
H_{abc} \xrightarrow{T_a} {f^a}_{bc} \xrightarrow{T_b} {Q_c}^{ab} \xrightarrow{T_c} R^{abc} \ .\label{Tdchain}
\eeq

This particular index structure actually relates geometric and non-geometric fluxes to the structure constants of the gauge algebras of gauged supergravities. Indeed, by first considering the Bianchi identities on the RR and NSNS fluxes, in absence of sources, and using T-duality, one can derive a set of constraints on the various fluxes. These conditions on NSNS fluxes can be identified \cite{stw05, dh05} with the Jacobi identities of a generalization of the gauge algebra of gauged supergravities. The latter is obtained by allowing, in addition to the usual fluxes in geometric situations \cite{km99, Derendinger:2004jn, df05, hr05}, the non-geometric fluxes as structure constants of the algebra.\footnote{The algebras involved generators of symmetries in the four-dimensional action, in particular generators of diffeomorphisms and of gauge transformations. These algebras were generalized to a T-duality covariant form: under the T-duality transformations of the chain \eqref{Tdchain}, the four types of fluxes in this chain then appeared as structure constants of these algebras.} We will come back to the conditions on the NSNS fluxes and the superpotential in section \ref{sec:Der}.

The relation to gauged supergravities was also discussed in the work on asymmetric orbifolds. Indeed, by considering the four-dimensional scalar potential, it was shown in \cite{dh02, dh05} that a Scherk-Schwarz reduction of an asymmetric orbifold could lead, in non-geometric situations, to a gauged supergravity. As a consequence, some gauged supergravities, which previously could not be lifted to a ten-dimensional supergravity geometric compactification, then finally found a higher dimensional, albeit non-geometric, interpretation \cite{dh05, dpst07}. 
 
Note that the few examples of ten-dimensional asymmetric orbifold descriptions are in essence stringy, and thus do not provide a relation between the four-dimensional non-geometric fluxes and the ill-defined supergravity fields in ten dimensions. The aim of this paper is to make a step further in establishing such a relation. This should clarify when the four-dimensional (super)potential provides a good effective description (as should be the case, for instance, in some toroidal compactifications, where the non-geometric situation is T-dual to a geometric one \cite{stw05, stw06}), and when its vacua can be lifted to string solutions \cite{stw06}. The following further development in non-geometry helped us to get a ten-dimensional understanding of these non-geometric fluxes.\\

An important aspect of non-geometry has been to give a geometrical description of the non-geometric configurations, at the cost of adding dimensions. As already discussed, using T-duality as transition functions, in addition to diffeomorphisms and gauge transformations, takes us away from the usual geometry of a smooth manifold. The resulting space has been called a T-fold \cite{h04}. Usual transition functions are elements of the structure
 group of the tangent bundle. In order for the T-duality group $O(n,n)$ to be the structure group of some bundle (of which diffeomorphisms and gauge transformations would appear as the ``geometric subgroup''), one needs to double the number of dimensions from $n$ to $2n$. A first way to do this is to double the number of
coordinates, which is known as doubled geometry (or, previously, the double formalism) and has motivations from string field theory \cite{h04}. A concrete example is the doubled torus: for a manifold consisting of a torus $T^n$ fibered over a base $\bbb$, one considers for each patch of $\bbb$ a new fiber $T^{2n}$. Only half of the coordinates and dimensions correspond to the physical space, and so on each patch one can define a polarization (or projection) towards this $T^n$. However, across patches, one can glue fields either with the standard diffeomorphisms and gauge transformations, or with a more involved T-duality element. The question
of non-geometry then arises globally: to get the physical space and fields, one glues together all the projected $T^n$. If this can be done smoothly (meaning 
if they are the same up to geometric transition functions), then one has a geometric configuration; otherwise it is non-geometric. The idea of the doubled geometry has then been used in several descriptions of non-geometry, along the lines of \cite{hr07, dpst07} for target space descriptions, and of \cite{h04, hm06, h06b} for sigma models.

Note that the T-fold, but also the initial idea of non-geometry, assume a local geometry; non-geometry appears as a global issue. Cases where we loose a local geometric description have also been considered; for instance in doubled geometry the fields would depend on the dual coordinates in a way that amounts to double not only the fiber, but also the base (see \cite{dh05, hr07} and references therein). Such non-geometric configurations should be obtained by performing T-dualities along non-isometry directions, in which case the standard Buscher rules cannot be applied.\footnote{\label{foot:TdR}There are some arguments in favour of such a generalized T-duality from string field theory, and proposals for such a transformation have been made \cite{eg95, h06a}. We also mentioned new rules for T-duality near \eqref{Tdchain}.} It can be argued, following in particular the T-duality chain \eqref{Tdchain} and the toroidal example, that such non-geometric situations would lead to (at least) an $R$-flux\footnote{Scherk-Schwarz reductions of duality twists, or asymmetric twists and shifts, could lead to such terms, providing then string realizations (at least classical ones) of non-geometric configurations of this kind. Examples of such non-geometries have been obtained from Kaluza-Klein monopoles and $N\!S5$-branes, see \cite{dh05, j11} and references therein.} while the locally geometric situation should give a $Q$-flux.\footnote{\label{foot:R}Note that the contrary is not true: having a $Q$- or an $R$-flux does not necessarily mean that we are facing a non-geometric situation. An example of a geometric compactification in a WZW model, providing an $R$-flux term, is given in \cite{dh05}. See also \cite{h09}, and the discussion on our rewriting in section \ref{sec:final}.} As we will discuss further, we will not consider $R$-fluxes in this paper, so we restrict ourselves to the T-fold case. The T-fold construction is related to the mathematical set-up of Generalized Complex Geometry (GCG) \cite{h02, g04}, which provides another way to consider a space of doubled dimension. This is the topic we now turn to.
 
\subsection{Non-geometry in Generalized Complex Geometry}\label{sec:nongeomGCG}

\subsubsection{Generalized tangent bundle}
In GCG, one considers the generalized tangent bundle $E$ which, for a manifold $\mmm$ of dimension $d$, consists of the fibration of the cotangent bundle over the tangent bundle. Locally, it is given by $T\mmm \oplus T^* \mmm$. Hence, the generalized tangent bundle also has doubled dimension, although the number of coordinates has not been doubled (in contrast to doubled geometry). A section of this bundle is given by
 a generalized vector $V$, which is a sum of a vector and a one-form: $V=v+\xi \in T\mmm \oplus T^* \mmm$. A natural metric $\eta$ coupling vectors and
 one-forms is given by the $2d \times 2d$ matrix
\beq
\eta=\begin{pmatrix} 0 & 1 \\ 1 & 0 \end{pmatrix} \ ,\qquad 
\frac{1}{2} \begin{pmatrix} \xi \\ v \end{pmatrix}^T \begin{pmatrix} 0 & 1 \\ 1 & 0 \end{pmatrix} \begin{pmatrix} \xi \\ v \end{pmatrix} = \xi_m v^m \ ,
\eeq
where the blocks are $d$-dimensional. The structure group of this bundle is then $O(d,d)$: it is the action on the generalized vectors which
preserves this metric and the inner product just defined. Indeed, for an action $V'=OV$ to preserve the inner product, one gets
\beq
O^T \eta O =\eta \Leftrightarrow O \in O(d,d) \ .
\eeq
Therefore, for each patch of $\mmm$, one can consider this new bundle $E$ with structure group $O(d,d,\mathbb{R})$. The T-duality group $O(n,n)$ (for $n$ isometries in $\mmm$) can be trivially embedded in $O(d,d)$. Analogously to the T-fold, one then uses the structure group of the bundle to glue fields from one patch to the other. This can be done with, for example, the geometric subgroup, or T-duality transformations. Given a metric $\hg$ and a $B$-field $\hB$ on $\mmm$, one can define the so-called generalized metric $\hhh$ as (see appendix \ref{ap:conv} for conventions)
\beq
\hhh=\begin{pmatrix} \hg-\hB \hg^{-1}\hB & \hB \hg^{-1} \\ -\hg^{-1} \hB & \hg^{-1}\end{pmatrix} \ . \label{genmet}
\eeq
One can show that $\hhh$ is of determinant one, and positive definite if $\hg$ is. Under the previous $O(d,d)$ action, $\hhh$ transforms as
\beq
\hhh'=O^T \hhh O \ .\label{eq:HO}
\eeq
This action exactly reproduces the T-duality transformations\footnote{More precisely, one should
consider the $n \times n$ blocks of the fields in the isometry directions, and act on them in the way just described with an $O(n,n)$ transformation in its
fundamental representation; here we consider for simplicity the whole $\hg$ and $\hB$, and a T-duality corresponds to embed $O(n,n)$ trivially in $O(d,d)$: the
action is non-trivial only along the isometry directions.} of the metric and the $B$-field \cite{gpr94}. Using this formalism is therefore very convenient
 for our purposes, and it actually leads to some insight on non-geometry, that we will now discuss. For completeness, let us add the T-duality
transformation of the dilaton
\beq
e^{-2\hp'} \sqrt{|\hg'|}=e^{-2\hp} \sqrt{|\hg|} \ ,\label{eq:Tddil}
\eeq
which leaves the measure of the NSNS action invariant.

\subsubsection{Generalized vielbeine, the \texorpdfstring{$\beta$}{H}-field, and non-geometric fluxes}

For a positive definite metric $g$ on $\mmm$, one can define vielbein matrices $e$ of coefficients $e^a_{\ m}$ as
\beq
g=e^T \id_d\ e \ .
\eeq
They are only defined up to $O(d)$ transformations: for $k \in O(d) \Leftrightarrow k^T \id_{d} k = \id_{d}$, $e$ and $k e$ give the
same metric. Correspondingly for the generalized metric $\hhh$ defined in \eqref{genmet}, one can introduce generalized vielbeine $\eee$
\beq \label{eq:HfromE}
\hhh= \eee^T \id_{2d}\ \eee \ ,
\eeq
also defined up to $O(2d)$ transformations: for $K \in O(2d)$, $\eee$ and $K\eee$ give the same generalized metric (see also footnote \ref{foot:sign}). There is however a
natural choice of basis (we label quantities in this basis with a hat), where the generalized vielbein takes the form
\beq
\heee=\begin{pmatrix} \he & 0 \\ -\he^{-T}\hB & \he^{-T} \end{pmatrix} \ . \label{genvielB}
\eeq
This generalized vielbein corresponds to the metric $\hg=\he^T \id_d\ \he$ and the $B$-field $\hB$ which appear in the generalized metric \eqref{genmet}. A generalized vielbein transforms under an $O(d,d)$ action as\footnote{\label{foot:busch}Note that performing a T-duality on a generalized vielbein can be more complicated: one sometimes has to act in addition with an $O(2d)$ transformation on the left. This is the case when one wants to reproduce the Buscher rules: to do so, one would act on $\hhh$ with an $O(d,d)$ element we denote $O_T$, which turns out to satisfy $O_T \in O(2d)$. The T-dual generalized vielbein is actually given by $\eee'=O_T \eee O_T$, see \cite{gmpw08, agmp10} for an illustration.}
\beq
\eee'=\eee O \ .
\eeq
In particular, $\heee$ transforms naturally under the geometric subgroup \cite{g04, gmpw08} consisting of diffeomorphisms and gauge transformations. However, different forms of generalized vielbeine  could equally well make sense on the generalized tangent bundle. Such a different form could be obtained after an $O(d,d)$
 transformation\footnote{The form of the generalized vielbein obtained after an $O(d,d)$ depends in particular on the vielbein $e$ before the
transformation. The $O(d)$ freedom on the choice of this initial vielbein is only a part of the $O(2d)$ freedom on the initial $\eee$, but it can be enough to give, after the $O(d,d)$ transformation, generalized vielbein of the form \eqref{genvielB} or \eqref{genvielbeta}.} on $\eee$, and/or an $O(2d)$ transformation
 on the left. Of particular interest for us is the following example (we denote quantities in this basis with a tilde)
\beq
\teee=\begin{pmatrix} \te & \te \tb \\ 0 & \te^{-T} \end{pmatrix} \ , \label{genvielbeta}
\eeq
where $\tb$ is a $d \times d$ antisymmetric matrix, that corresponds to an antisymmetric  bivector $\tb^{mn}$ when we put back indices. This object will play a key role for us: indeed, it has been argued \cite{gs06, e06, mpt07, gs07, gmpw08} that the appearance of such a $\beta$ block can be a sign of non-geometry. Before giving details on this point, let us mention that such a  bivector first appeared in a particular $O(d,d)$ transformation named the $\beta$-transformation \cite{g04}. The latter was used at first \cite{mpz06} in the context of AdS/CFT, to describe in particular the Lunin-Maldacena solution \cite{lm05}.

As discussed previously, the generalized vielbein $\teee$ in which $\tb$ appears is not the one which transforms naturally\footnote{\label{foot:gaugetb}One could still imagine a gauge transformation on $\tb$ instead of $\hB$ and so propose an alternative geometric subgroup. It seems however that such a gauge transformation would be difficult to define because of cocycle conditions \cite{gmpw08}.} under the geometric subgroup: that would be $\heee$. Hence, even if one can make sense of a vielbein like $\teee$ on the generalized tangent bundle, coming back to the standard geometry amounts to use $\heee$. In the
T-fold language, we could say that the polarization makes us consider a vielbein of the form $\heee$ on each patch. Locally, it is always possible to go from a generalized vielbein of the form $\teee$ to one like $\heee$, by using an $O(2d)$ transformation (more precisely a particular $O(d)\times O(d)$ transformation). However, if $\tb$ is ill-defined with respect to $\mmm$ coordinates, the $O(d) \times O(d)$ transformation might not be single-valued, and so cannot be performed globally. This means that one cannot go back to the standard geometric description globally on $\mmm$, and is a sign of non-geometry \cite{gmpw08} (see also the last appendix of \cite{agmp10} for an illustration).

In addition to the above argument, a relation between $\tb$ and non-geometry also results from algebraic considerations. The action of the Courant bracket, the natural bracket in GCG, on various components of generalized vielbeine has been analyzed in \cite{gmpw08}. The resulting algebra appears as the natural equivalent in GCG of the gauge algebras of gauged supergravity. As discussed previously, some of the structure constants of these gauge algebras  correspond to the non-geometric fluxes. By looking at the action of the Courant bracket, using different bases for the generalized vielbein (in particular $\heee$ or $\teee$), and also by applying T-dualities on some concrete examples (like the toroidal example), the structure constants of the corresponding algebra were related to $\tb$. More precisely, the following expressions were obtained for the non-geometric fluxes with flat space indices
\beq
{Q_c}^{ab} = \del_c \tb^{ab} - 2 \tb^{d [a} \tf^{b]}_{\ cd} \ , \ R^{abc}=-2 \tb^{d [a} \del_d \tb^{b]c} + \tb^{ad} \tb^{be} \tf^c_{\ de} \ .\label{nongeoflux}
\eeq
The structure constants $\tf^{a}_{\ bc}$ are defined as in \eqref{struct} with respect to vielbeine $\te$ of \eqref{genvielbeta}. Note that taking $\tb^{ab}$ as a tensor, we obtain 
\beq
{Q_c}^{ab} = \te_c^{\ m} \te^a_{\ n} \te^b_{\ p}\del_m \tb^{np} + 2 \tb^{nm} \te_c^{\ p} \te^{[a}_{\ \ n} \del_m \te^{b]}_{\ p} \ ,
\label{eq:Q}
\eeq
where the second term vanishes with our assumption \eqref{eq:intassume}. Again, the expressions in \eqref{nongeoflux} relate a non-trivial $\tb$ to non-geometry.\\

In \cite{h08, h09}, string world-sheet analyses were used to derive a more general form of \eqref{nongeoflux}, for a set-up with both $B$ and $\beta$ non-zero. In \cite{h08}, by considering world-sheet Hamiltonians, and proposing a coupling of a bivector to the bosonic string, a particular bracket and associated charges are derived. The bracket is the Roytenberg bracket which generalizes the Courant bracket, and its charges are related to non-geometric fluxes. Expressions for those are given in terms of the bivector that we call $\beta$ here. The Hamiltonians considered in \cite{h08} found deeper roots in \cite{h09}, where a first order action reproducing them is proposed. This first order action considers again a coupling to a bivector, in addition to a two-form and a metric, and is related to the NSNS string. It can also be related to a three-dimensional membrane action, where the whole set of (non)-geometric fluxes appear as couplings. The expressions for these fluxes are reobtained, in particular
\beq
{Q_c}^{ab} = \del_c \beta^{ab} - 2 \beta^{d [a} f^{b]}_{\ cd} + H_{cde} \beta^{da} \beta^{eb} \ .\label{eq:QH}
\eeq
Given this formula is valid in presence of a $B$-field, this situation should not correspond to either the hatted or tilded bases considered so far, so we drop these notations here. We will return to such a situation in section \ref{sec:disc}.\\

At this stage, all these expressions of the non-geometric fluxes remain somewhat formal, and they were derived from rather different perspectives. Their interpretation in ten dimensions is not clear, and no higher dimensional action including them has been proposed. Thus no dimensional reduction has really been made, even if a relation to the four-dimensional superpotential has been established, and we discuss  the latter in the following. Given the rewriting done in this paper, the ten-dimensional understanding of these formulas, and the relation to four dimensions, should be clearer.

\subsubsection{Generalized covariant derivative and superpotential}\label{sec:Der}

We recall from section \ref{sec:rev} that the Jacobi identities of the gauge algebras lead to a set of constraints on the geometric and non-geometric NSNS fluxes, and that those can also be obtained by applying T-duality on the Bianchi identity $\d H=0$. In a geometric setting, one often encounters the twisted exterior derivative $\d - H$. Asking for its nilpotency results exactly in the previous $H$-flux Bianchi identity. Inspired by this situation, one can define a generalized covariant derivative $\ddd$, see \eqref{eq:gencovder},  where not only $H$ and the geometric fluxes enter (the latter from the exterior derivative), but also the $Q$- and $R$-fluxes. $\ddd$ is then covariant under T-duality. It turns out \cite{stw06, irw07} that requiring its nilpotency results in the Jacobi identities of the gauge algebra, together with the unimodularity condition $f^a_{\ ab}=0$, and the condition
\beq
{Q_a}^{ab}=0 \ .\label{eq:traceQ}
\eeq
We come back to the relation between $\ddd$ and constraints on the fluxes in section \ref{sec:bi}.

A deeper explanation for the generalized covariant derivative $\ddd$ can be found in GCG. Indeed, one can define in GCG a derivative in terms of a generalized spin connection. The latter is related to the structure constants one obtains from the Courant bracket, when acting on the generalized vielbeine. As explained previously, these structure constants should be given by the (non)-geometric fluxes, and so the derivative defined in GCG should correspond to the generalized covariant derivative $\ddd$ (see \cite{gmpw08} and references therein). This brings more credit to the expressions \eqref{nongeoflux} of the non-geometric fluxes.

For a geometric compactification on a background admitting an $SU(3)\times SU(3)$ structure, the four-dimensional superpotential can be written in terms of an integral over GCG objects \cite{glw05, bg06, glw06}. It involves the twisted derivative $\d -H$. The covariant derivative $\ddd$ then allows a natural generalization of this expression, that includes non-geometric fluxes. The generalized formula was shown \cite{glw06, mpt07} to reproduce the four-dimensional superpotentials previously mentioned \cite{stw05, acfi06, stw06}, after compactification (see also \cite{irw07, rw07}).

Let us note that in this process, one has to integrate over internal fields. To justify an integration over naively ill-defined fields, the covariant derivative $\ddd$ plays an important role. When going from one patch to the other, the fields of a non-geometric configuration can be glued via T-duality transformations. The same goes for the integrand of the superpotential, and this is the only way to make it globally defined. This would a priori make the integration rather involved. However, by expressing the integrand in terms of the covariant derivative, it turns out to be invariant under a T-duality transformation \cite{mpt07}, therefore it is single-valued and one can integrate.\footnote{The K\"ahler potential is also invariant \cite{glw05}, so the scalar potential is well-defined.} A drawback of this argument is that it requires to define a $Q$-flux in ten dimensions (within $\ddd$), which is not realized so far. This paper should help in this direction. Then, we return to the question of the integration in section \ref{sec:totder} (see also footnote \ref{foot:Der}).

\section{Rewriting of the NSNS Lagrangian}\label{sec:computation}

In this section, we come back in detail to the rewriting of the NSNS Lagrangian \eqref{eq:intlag}. From the previous sections, we first repeat some arguments to motivate such a rewriting. We also discuss the assumption \eqref{eq:intassume} used in our computation. Then we give the details of the derivation of the equality \eqref{eq:intlag}, and finally comment on this result.

\subsection{Setting the stage}

\subsubsection{Motivation}

Suppose we perform
 T-dualities on a geometric solution, and end up with a non-geometric solution. In four dimensions, performing these T-dualities leads to new terms in
 the potential, and in the gauge algebra, corresponding to non-geometric fluxes. On the contrary, in ten dimensions, we do not see any new degree of
 freedom appearing: we only deal with NSNS fields, which are globally ill-defined. However, in GCG, it is argued that non-geometry can be seen by looking at
 generalized vielbeine of the form $\teee$ \eqref{genvielbeta} where an antisymmetric bivector $\tb^{mn}$ appears. In addition, the ill-defined metric
 $\hg$ and $B$-field $\hB$, present in the generalized metric $\hhh$ and in the generalized vielbein $\heee$ \eqref{genvielB}, can be reexpressed in terms
 of this $\tb$ and a new metric $\tg$. Indeed, one should keep in mind that even if the generalized vielbeine can be different, the generalized metric should be the same. So one
has the relation \eqref{eq:intequality}, that we give again here:
\beq
\hhh=\begin{pmatrix} \hg-\hB \hg^{-1}\hB & \hB \hg^{-1} \\ -\hg^{-1} \hB & \hg^{-1}\end{pmatrix} = \heee^T \id_{2d}\ \heee = \teee^T \id_{2d}\ \teee 
= \begin{pmatrix} \tg & \tg\tb \\ -\tb \tg & \tg^{-1}-\tb \tg \tb \end{pmatrix} \ . \label{equality}
\eeq
This equality provides us with relations between the hatted and the tilded fields, allowing us to perform a change of variable in the NSNS Lagrangian. We can then benefit from the insight of GCG on the role played by $\tb$ in non-geometric situations. In particular, $\tb$ is related to non-geometric fluxes via the
expressions \eqref{nongeoflux}. So one can guess that performing the change of variable in the NSNS Lagrangian would make the non-geometric fluxes appear. We
 show in the following that this is actually true, at least for a subcase where we use the simplifying assumption \eqref{eq:intassume}.

\subsubsection{Definitions}\label{sec:def}

Let us now turn to this change of variables more explicitly. The equality \eqref{equality} gives the following relations\footnote{\label{foot:exphB}To get the expression for $\hB$, one can first notice that
\beq
\hB \hg^{-1}= \tg \tb = \frac{1}{2} \tg\left((\tg^{-1}+ \tb) - (\tg^{-1}- \tb)  \right) \ .\label{firstB}
\eeq
Note as well that the symmetry or antisymmetry of $\hg$ and $\hB$ is automatic in this parametrization. One can also check that while \eqref{equality} first
gives the two expressions \eqref{ghat} and \eqref{Bhat} of $\hg$ and $\hB$, the third constraint we get from it is then automatically satisfied. Let us note that the left equality of \eqref{firstB} and the first line of \eqref{gtilde} appeared already in \cite{h08, gmpw08}.}
\bea
\hg &= \left(\tg^{-1} - \tb \tg \tb \right)^{-1} \nn\\
&= (\tg^{-1}+ \epsilon \tb)^{-1} \tg^{-1} (\tg^{-1}- \epsilon \tb)^{-1} \ , \ \epsilon=\pm1 \nn \\
&= \frac{1}{2}\left((\tg^{-1}- \tb)^{-1} + (\tg^{-1}+ \tb)^{-1} \right) \label{ghat}\\
\nn\\
\hB&= (\tg^{-1}+ \varepsilon \tb)^{-1} \tb (\tg^{-1}- \varepsilon \tb)^{-1} \ , \ \varepsilon=\pm1 \nn \\
&= \frac{1}{2}\left((\tg^{-1}- \tb)^{-1} - (\tg^{-1}+ \tb)^{-1} \right) \ , \label{Bhat}
\eea
and the converse relations
\bea
\tg &= \hg - \hB \hg^{-1} \hB \nn\\
&= (\hg+ \epsilon \hB) \hg^{-1} (\hg- \epsilon \hB) \ , \ \epsilon=\pm1 \nn \\
&= \frac{1}{2}\left((\hg - \hB)^{-1} + (\hg + \hB)^{-1} \right)^{-1} \label{gtilde}\\
\nn\\
\tb&= (\hg + \varepsilon \hB)^{-1} \hB (\hg - \varepsilon \hB)^{-1} \ , \ \varepsilon=\pm1 \nn \\
&= \frac{1}{2}\left((\hg - \hB)^{-1} - (\hg + \hB)^{-1} \right) \ . \label{betatilde}
\eea
 
For later convenience, we introduce the notation
\beq
G_{\pm}^{mn}=\tg^{mn}\pm\tb^{mn} \ , \label{eq:defG}
\eeq
where one can notice that
\beq
G_{\pm}^T=G_{\mp} \ .
\eeq
This property will allow us in the following to use mainly $G_+$, that we will denote for simplicity as $G=G_+$. The definition \eqref{eq:defG} allows us
 to rewrite \eqref{ghat} and \eqref{Bhat} as
\beq
\hg_{mn}=(G^{-1}_{\pm})_{mk} \tg^{kp} (G^{-1}_{\pm})_{np}\ , \ \ \hg^{mn}=G_{\pm}^{mk} \tg_{kp} G_{\pm}^{np}\ ,\ 
 \ \hB_{mn} = (G^{-1}_{\pm})_{mk} \tb^{kp} (G^{-1}_{\mp})_{pn} \ . \label{eq:convrewr}
\eeq
Finally, we also define a dilaton $\tp$, such that the measure remains invariant (see appendix \ref{ap:conv} for our conventions)
\beq
e^{-2\tp} \sqrt{|\tg|}=e^{-2\hp} \sqrt{|\hg|} \ . \label{dilmeasure}
\eeq
This definition is inspired by T-duality, which leaves the measure invariant (see \eqref{eq:Tddil}). After a short computation, we get\footnote{\label{foot:detg}From \eqref{eq:convrewr}, we get $(\det (\hg))^{-1}= (\det (G))^2 \det(\tg)$ which implies that $\det (\hg)$ and $\det (\tg)$ have the same sign. In their ratio, we can then drop the absolute value, and we use \eqref{eq:lndet} to get the second equality of \eqref{tp}.}
\beq
e^{-2\tp} =e^{-2\hp} |\id_d-\tb \tg \tb \tg|^{-\frac{1}{2}} \quad \Leftrightarrow \quad \hp = \tp - \frac{1}{4}{\rm tr}(\ln(\id_d-\tb \tg \tb \tg)) \ . \label{tp}
\eeq

We can now use the expressions \eqref{ghat}, \eqref{Bhat} and \eqref{tp} to rewrite the NSNS Lagrangian in terms of the variables $\tg$, $\tb$ and
$\tp$. This computation turns out to be rather involved, and so we make use of a simplifying assumption, that we now discuss. The computation
 without using the assumption is work in progress \cite{a11b}.

\subsection{Discussion of the assumption}

To derive the rewriting of the NSNS Lagrangian \eqref{eq:intlag}, we use the assumption \eqref{eq:intassume} that we recall here:
\beq
\tb^{km}\partial_m\cdot=0\ , \label{eq:assume}
\eeq
where the dot is a placeholder for any field. Since the fields in one basis are combinations of those in the other basis, thanks to the relations
 \eqref{ghat}, \eqref{Bhat} and \eqref{tp}, we can simply say that the dot stands for $\hg$, $\hB$ and $\hp$. We refine the statement by saying
 that the dot can also stand for vielbeine. This allows to apply \eqref{eq:assume} to fields with both curved or flat indices.

One advantage of this assumption is to simplify the computation, which is otherwise rather involved. The difficulty comes from the combination $G^{-1}=(\tg^{-1}+\tb)^{-1}$ which appears quite often, and cannot be simplified or developed in any manner (see in the next section the number of terms involving it). The assumed property \eqref{eq:assume} of $\tb$ helps us to simplify some expressions where $G^{-1}$ appears, as we will see below. Another reason to use \eqref{eq:assume} is that the expressions of the non-geometric fluxes \eqref{nongeoflux} reduce to
\beq
{Q_m}^{np}=\del_m \tb^{np} \ , \ R^{mnp}=0 \ . \label{eq:asnongeoflux}
\eeq
This can be seen using \eqref{eq:Q} and \eqref{struct}, assuming $\tb$, $Q$ and $R$ to be tensors. Even if the situation given by \eqref{eq:asnongeoflux} is not the most general, it prevents us from considering the conceptual issues arising in the presence of an $R$-flux (see previous
 section). Making this (simpler) $Q$-flux appear in the NSNS Lagrangian is minimal but sufficient to show the interest of the rewriting, and we restrict
ourselves to that particular set-up in this paper.

An important observation is that the assumption \eqref{eq:assume} holds in the toroidal example of appendix \ref{ap:toyex}, which is a standard example of non-geometric situations obtained by T-duality. Note that our computation does not rely on the use of T-duality, neither as a (non-geometric) solution generating technique, nor as the string symmetry used to make fields single-valued. Nevertheless, the validity of this assumption is inspired by T-duality, as we now explain. The derivation of the Buscher rules \cite{b87, b88} requires the NSNS fields to be independent of some of the coordinates. Locally, this is equivalent to having some isometries. The global aspects can complicate things (see \cite{gmpw08} and references therein for a discussion on topological aspects and T-duality), but for simplicity we will talk here of an isometry direction when all the fields are independent of the corresponding coordinate. We then define the fiber $\fff$ as the subspace along the isometry directions, and the base space $\bbb$ along the other directions. This is in agreement with the definitions of $\fff$ and $\bbb$ in section \ref{sec:rev}. For instance, in the toroidal example, $\fff$ is along $\del_{x}$ and $\del_{y}$, while $\bbb$ is along $\del_{z}$. Given these definitions, the assumption \eqref{eq:assume} is satisfied if $\tb$ is purely\footnote{$\tb$ being purely along $\fff$ implies the assumption, but the converse is strictly speaking not true, since the index $m$ in
 \eqref{eq:assume} is summed.} along $\fff$, meaning any of its components with a leg along $\bbb$ is zero.

When is this property of $\tb$ valid? From \eqref{betatilde}, we can see this property is verified if $\hB$ is purely along $\fff$, together with the
 manifold having a product structure (i.e. the metric $\hg$ is block diagonal along $\fff$ and $\bbb$). We recall this is the case in any frame in the toroidal
 example, and so the assumption is automatic there.\footnote{We mentioned in section \ref{sec:nongeomGCG} that in GCG, a $\tb$ could appear in a generalized
vielbein $\eee'$ after applying T-dualities along $\fff$ on an initial generalized vielbein $\eee$ of the form $\heee$ \eqref{genvielB}. The (Buscher) T-dual of $\eee$ is
 given by $O_T \eee O_T$ with $O_T \in O(d,d)$, see footnote \ref{foot:busch}. If one obtains after T-duality an $\eee'$ of the form $\teee$
 \eqref{genvielbeta}, one can show that $\tb$ is purely along $\fff$ provided the initial vielbein has the same properties we just discussed: the initial
 $B$-field is purely along $\fff$, and the initial vielbein is block diagonal along $\fff$ and $\bbb$.}\\

The assumption \eqref{eq:assume} implies the simplifications
\beq
G_{\pm}^{nk}\partial_k\cdot=\tg^{nk}\partial_k\cdot \quad , \quad \tg_{mn}G_{\pm}^{nk}\partial_k\cdot=\partial_m\cdot \quad ,\quad
 \left(G^{-1}_\pm\right)_{mn}\tg^{nk}\partial_k\cdot=\partial_m\cdot \quad ,\label{eq:simplif}
\eeq
that we will use. Furthermore, it leads to two other properties, and in the following, we will keep track of their use. From \eqref{eq:assume}, we can 
first deduce $\partial_k \left( \tb^{nm}\partial_m\cdot \right)=0$, hence
\beq
\partial_k \tb^{nm}\partial_m\cdot=0 \ . \label{as1bis}
\eeq
Similarly, we also consider
\beq
\partial_k \tb^{kn}=0 \ . \label{as1ter}
\eeq
This second property can be obtained from an integration by parts on any Lagrangian term involving \eqref{eq:assume}.\footnote{For $F_n$ and $f$ some combination of fields, one has
\beq
e^{-2\tp} \sqrt{|\tg|}\ F_{n} \tb^{nk} \del_k f = \del_k(\tb^{nk} \dots) 
- \left( \del_k \left(e^{-2\tp} \sqrt{|\tg|}\ F_{n} \right) \tb^{nk} + e^{-2\tp} \sqrt{|\tg|}\ F_{n} \del_k (\tb^{nk}) \right) f \ . \nn
\eeq
The left-hand side vanishes, out of \eqref{eq:assume}, and so does the second term on the right-hand side. Therefore, discarding the total derivative term, we conclude that \eqref{as1ter} holds. One could argue though that what is inside the total derivative, is a priori 
not globally defined. As a consequence, this term should not vanish (see section \ref{sec:totder}). This reasoning would be correct if the derivative index were independent of what was acted on, but it is not the case here. If all field components (including $\tb$ itself) depend on a certain set of coordinates corresponding to directions $q$, then the assumption \eqref{eq:assume} essentially amounts to having the components $\tb^{pq}$ ($\forall p$) vanishing. Therefore, in the sum on $k$ given by $\del_k(\tb^{nk} \dots) $, either the derivative gives zero, or the component $\tb^{nk}$ is zero. So this total derivative term can be discarded because it vanishes. The same reasoning could in fact be applied on $\partial_k \tb^{kn}$ itself.}  It also corresponds to ${Q_k}^{kn}=0$, which is a known constraint \eqref{eq:traceQ} that should be imposed on the $Q$-flux.

\subsection{Rewriting step by step}

We now explain the rewriting of each term in the NSNS Lagrangian, when expressing $\hg$, $\hB$ and $\hp$ in terms of $\tg$, $\tb$ and $\tp$,
and making use of the assumption just discussed. The technical details are given in appendix \ref{ap:comp}.

\subsubsection{The Ricci scalar}\label{sec:Ricci}

In the appendix, we give general formulas related to the Ricci scalar. For both $\hg$ and $\tg$, we use the Levi-Civita connection.
 In particular, we get from \eqref{tR} the Ricci scalar for $\tg$
\bea
\tR &= \tg^{lm}\tg^{ku} \del_k \del_m \tg_{lu} - \tg^{lu}\tg^{km} \del_k\del_m \tg_{lu} \vs \label{Rtilde}\\
&\quad +\tfrac{1}{2} \del_m \tg_{ln} \del_k \tg_{pu} \left( 2 \tg^{kl} \tg^{mn} \tg^{pu} - \tfrac{1}{2} \tg^{km} \tg^{ln} \tg^{pu}\right.\nn\\
&\quad \left. + \tfrac{3}{2} \tg^{km} \tg^{np} \tg^{lu} - \tg^{mp} \tg^{kn} \tg^{lu} - 2 \tg^{mn} \tg^{kp} \tg^{lu} \right) \ .\nn
\eea
The computation of $\hR$ in terms of $\tg$ and $\tb$ is rather involved and we detail it in the appendix. We obtain a first expression using the assumption
\eqref{eq:assume} and the simplifications \eqref{eq:simplif}, and then slightly rewrite it for later convenience as
\bea
\hR-\tR=& - \partial_{k}\tg_{su} \partial_{m}\tg_{pq} \left(2\tg^{km}\tg^{uq}\tg^{ps}+2\tg^{pq}\tg^{ks}\tg^{mu}+\frac{1}{2}\tg^{uq}\tg^{sm}\tg^{kp}\right) \label{eq:R-Ras}\\
& -\tg_{pq} \partial_{k}\tb^{pk} \partial_{m}\tb^{qm} -\frac{1}{2}\tg_{pq} \partial_{k}\tb^{qm} \partial_{m}\tb^{pk}  \nn\\
& +2\tg^{km}\tg^{pq} \partial_{k}\partial_{m}\tg_{pq} +2\tg^{km}\iG_{pq} \partial_{k}\partial_{m}G^{qp} \nn\\
& + \partial_{m}G^{vl} \Big( - 2 \tg^{mr} \tg^{ks} \iG_{lv} \del_k \tg_{rs} -\tg^{rs}\tg^{km} \iG_{lv} \partial_{k}\tg_{rs} \nn\\
& \quad +\tg^{ms}\tg^{ru}\iG_{lu} \partial_{v}\tg_{rs} -\tg^{km}\tg^{rs}\iG_{ls} \partial_{k}\tg_{vr} \Big) \nn\\
& + \partial_{m}G^{vl} \Big(\iG_{lq} \partial_{v}G^{qm} +\frac{1}{2} \hg_{lq} \partial_{v} G^{mq} \Big) \nn\\
& - \partial_{m}G^{vl}  \partial_{k}G^{ps}  \frac{1}{2} \tg^{km} \Big(2 \iG_{lv}\iG_{sp} + 5 \iG_{sv}\iG_{lp}  + \hg_{sl}\tg_{pv} \Big)\ .\nn
\eea
Note that a few $\hg_{mn}$ are kept to shorten notations: they should simply be replaced in terms of $\tg$ and $\iG$ using
 \eqref{eq:convrewr}.

Let us make a few remarks on \eqref{eq:R-Ras}. The assumption has helped to simplify some complicated terms, but a few terms containing $\iG$ remain. As mentioned in the previous section, these terms cannot be simplified further, which is what makes this computation difficult. Fortunately, all the $\iG$ terms will in the end cancel against other terms in the Lagrangian, to leave us only with the non-geometric $Q$-flux. Let us note also that the third line of \eqref{eq:R-Ras} contains second order derivative terms. These cannot cancel against any other since the other terms of the Lagrangian only contain first order derivatives. We thus need to integrate these terms by parts, which will give us in particular a total derivative term, that is discussed in more detail in section \ref{sec:totder}. The integration by parts will also give rise to terms with a derivative of the dilaton. Therefore, we prefer to turn to the dilaton kinetic terms, before treating these second order derivative terms.

\subsubsection{The dilaton terms}

Starting from the definition \eqref{tp} of $\hp$ in terms of $\tp$, we compute in the appendix the difference between the two dilaton terms (each squared
with its corresponding metric, $\hg$ or $\tg$) and get
\bea
|\d\hp|^2-|d\tp|^2 &= \tfrac{1}{4} \tg^{km} \tg^{pq} \tg^{uv} \partial_m \tg_{pq} \partial_k \tg_{uv}  \label{eq:dhpminusdtp}\vs\\
&\quad + \tfrac{1}{2} \tg^{km} \tg^{pq} \iG_{uv} \partial_m \tg_{pq} \partial_k G^{vu} \vs\nonumber\\
&\quad + \tfrac{1}{4} \tg^{km} \iG_{pl} \iG_{uv} \partial_m G^{lp} \partial_k G^{vu} \nonumber\\
&\quad - \tg^{km} \tg^{pq} \partial_k \tg_{pq} \partial_m \tp \nonumber\\
&\quad - \tg^{km} \iG_{pq} \partial_{k} G^{qp} \partial_m \tp \ , \nn
\eea
where we used the assumption \eqref{eq:assume} and the simplifications \eqref{eq:simplif}.

\subsubsection{Second order derivative terms and integration by parts}

We come back to the second order derivative terms appearing in \eqref{eq:R-Ras} and the integration
by parts. For combinations of fields $f$ and $F^{km}$ (of indices $k, m$), we have
\bea
\int \d^D x\ e^{-2\tp} \sqrt{|\tg|}\ F^{km} \del_k \del_m f =& \int \d^D x\ \del_k(\dots) \\
+ & \int \d^D x\ e^{-2\tp} \sqrt{|\tg|}\ \Bigg( \left( 2\del_k \tp -\frac{1}{2} \tg^{pq} \del_k \tg_{pq} \right) F^{km} -\del_k F^{km} \Bigg) \del_m f \ ,
\eea
where we used the tilded measure for convenience, but note that the hatted and tilded measures are the same \eqref{dilmeasure}. Applied to the third line of
 \eqref{eq:R-Ras} one gets
\bea
2 \tg^{km} \tg^{pq} & \del_k \del_m \tg_{pq} + 2 \tg^{km} \iG_{pq} \del_k \del_m G^{qp} =\frac{1}{e^{-2\tp} \sqrt{|\tg|}} \partial_k (\dots)  \label{eq:intbyparts} \vs \\
& + 4 \tg^{km} \del_k \tp \left( \tg^{pq} \del_m \tg_{pq} + \iG_{pq} \del_m G^{qp} \right) \nn \\
& + \del_k \tg_{uv} \del_m \tg_{pq} \left( 2 \tg^{pq} \tg^{mu} \tg^{kv} + \tg^{km} \left( 2 \tg^{pu} \tg^{vq} - \tg^{uv} \tg^{pq} \right) \right) \nn \\
& + \iG_{pq} \del_k \tg_{uv} \del_m G^{qp} \left( 2 \tg^{mu} \tg^{kv} - \tg^{km} \tg^{uv} \right) \nn \\
& + 2 \tg^{km} \iG_{pu} \iG_{vq} \del_k G^{uv} \del_m G^{qp} \vs \nn\ ,
\eea
where the total derivative is given by
\beq
\partial_k \left( e^{-2\tp} \sqrt{|\tg|}\ 2 \tg^{km} \left( \tg^{pq} \partial_m \tg_{pq} + (G^{-1})_{pq} \partial_m G^{qp} \right) \right)\ . \label{eq:totalderivative}
\eeq
This can be expressed purely in terms of hatted fields, or tilded fields, using formulas \eqref{ghat}, \eqref{Bhat}, or their converse. 

Note that some terms appearing in \eqref{eq:intbyparts} exactly cancel terms in \eqref{eq:dhpminusdtp}, in particular those which contain a derivative of $\tp$. Replacing the second order derivative terms in \eqref{eq:R-Ras} by \eqref{eq:intbyparts}, adding \eqref{eq:dhpminusdtp}, and using (for the first time) the two other
 properties \eqref{as1bis} and \eqref{as1ter} derived from the assumption, one finally obtains
\bea
\hR -\tR &+ 4 (|\d \hp|^2- |\d \tp|^2) - \frac{\del(\dots)}{e^{-2\tp} \sqrt{|\tg|}} \label{Rphidas}\\
=& \frac{1}{2}  \tg^{ku} \tg^{mq} \tg^{vl} \tg^{sp} (\hg_{sl}- \tg_{sl}) \del_m \tg_{uv} \del_k \tg_{pq} \nn\\
& + \tg^{km} \del_m G^{pl} \left(\tg_{pq} \iG_{lr} \del_k \tg^{rq} - \frac{1}{2} \left(\hg_{lq} \tg_{pr} + \iG_{qp} \iG_{lr} \right) \del_k G^{rq} \right) \ ,\nn
\eea
where the total derivative is given by \eqref{eq:totalderivative}. We can  easily verify that the right-hand side of \eqref{Rphidas} vanishes for $\tb=0$.

\subsubsection{The \texorpdfstring{$H$}{H}-flux term}

The NSNS $H$-flux is given by the three-form $\hH=\d \hB$, where the coefficient of the two-form $\hB$ is given in \eqref{eq:convrewr}. Following our
conventions on forms in appendix \ref{ap:conv}, we compute in  appendix \ref{ap:comp} the $|\hH|^2$ term in the NSNS Lagrangian. Using the assumption
 \eqref{eq:assume}, together with \eqref{eq:simplif} and \eqref{as1bis} (but not \eqref{as1ter}), we obtain an expression for $|\hH|^2$. This is then
 rewritten to obtain an expression more easily comparable with \eqref{Rphidas}:
\bea
|\hH|^2 &= \tfrac{1}{2} \tg_{p_1 p_2} \tg_{q_1 q_2} \tg^{s_1 s_2} \del_{s_1} \tb^{p_1 q_1}  \del_{s_2} \tb^{p_2 q_2} \label{eq:Hsquared}\vs\\
&\quad + (\hg_{q_1 q_2} - \tg_{q_1 q_2}) \del_{p_2} \tg^{p_1 q_1} \del_{p_1} \tg^{p_2 q_2} \vs\nn\\
&\quad +  \tg^{s_1 s_2} \del_{s_1} G^{p_1 q_1} \Bigg(2\tg_{p_1 p_2}  (G^{-1})_{q_1 q_2} \del_{s_2} \tg^{p_2 q_2} \nn\\
&\quad\;\; - \left( \tg_{p_1 p_2} \hg_{q_1 q_2} + (G^{-1})_{q_2 p_1} (G^{-1})_{q_1 p_2} \right)  \del_{s_2} G^{p_2 q_2}\Bigg) \ .\nn
\eea

\subsubsection{Final result and comments}\label{sec:final}

Combining all the expressions computed, namely \eqref{Rphidas} and \eqref{eq:Hsquared}, most of the terms drop out, and we are left with
\beq
\hR -\tR + 4 (|\d \hp|^2- |\d \tp|^2) - \frac{\del(\dots)}{e^{-2\tp} \sqrt{|\tg|}} - \frac{1}{2} |\hH|^2 = -\frac{1}{4} \tg_{p_1 p_2} \tg_{q_1 q_2} \tg^{s_1 s_2} \del_{s_1} \tb^{p_1 q_1}  \del_{s_2} \tb^{p_2 q_2} \ .
\eeq
Using \eqref{eq:asnongeoflux}, \eqref{eq:Q2R2} and \eqref{dilmeasure}, we obtain the rewriting of the NSNS Lagrangian \eqref{eq:intlag} discussed in the introduction 
\beq
e^{-2\hp} \sqrt{|\hg|} \left(\hR + 4|\d \hp|^2 - \frac{1}{2} |\hH|^2 \right) =e^{-2\tp} \sqrt{|\tg|} \left(\tR + 4|\d \tp|^2 
- \frac{1}{2} |Q|^2 \right) 
+ \del (\dots) \label{eq:equa} \ ,
\eeq
where the total derivative term is given in \eqref{eq:totalderivative}.

The final result may at first look surprising, because it was derived  without specifying anything about the fields (except the assumption \eqref{eq:intassume}). In particular, it was not mentioned whether the field configuration was geometric or not, and nevertheless the result indicates it is always possible to obtain a non-geometric $Q$-flux. An explanation for this has actually already been given, for example in section \ref{sec:nongeomGCG}: locally, it is always possible to go from the hatted basis to the tilded basis via an $O(2d)$ transformation on the generalized vielbein; the question of non-geometry only arises as a global issue \cite{gmpw08}.\footnote{This is related to having a topological obstruction which prevents from performing T-duality, see related discussion in \cite{h06a, gmpw08} and references therein.} Similarly here, the Lagrangian is only a local quantity and can always be rewritten in a form that contains a $Q$-flux term. However, once we integrate the Lagrangian to obtain an action, the global aspects become important, and we really face the question of non-geometry. This will be discussed further in section \ref{sec:totder}.

\section{Aspects of the rewritten action}
\label{sec:disc}

\subsection{Global aspects: preferred bases and total derivative term}\label{sec:totder}

In section \ref{sec:computation}, we derived the equality \eqref{eq:equa} of the Lagrangians $\hL$ and $\tL$, up to a total derivative term. In this section, we are interested in the associated actions, so we focus on the global aspects. In particular, the global properties of the fields, and consequently of the Lagrangians, will play an important role. For instance, the NSNS fields in a non-geometric configuration are globally ill-defined, since they are not single-valued when using only standard geometric transition functions. As a consequence, it is difficult to consider an action over these fields, because the usual integration cannot be performed. The global aspects of the fields are also important in the total derivative term. For globally well-defined fields, such a term is discarded, since its integral over a manifold without boundary vanishes. 
This is not the case for a non-geometric configuration, so the fate of the total derivative term in \eqref{eq:equa} does matter. In particular, whether it integrates to zero tells us if we get the same action in the two bases. This is important, since we would like to know whether we can trade the NSNS action for the one obtained by integrating $\tL$.\\

The toroidal example presented in appendix \ref{ap:toyex} provides a good illustration for the discussion of these various questions, so let us first comment on it. In the T-duality frames A and B, the hatted fields describe a geometric configuration (a torus together with a $B$-field in frame A, or a twisted torus in frame B). The fields are well-defined and so is the NSNS Lagrangian $\hL$. What about the tilded fields and $\tL$? In frame B, the two bases are identical, and the two Lagrangians are the same and well-defined. Frame A is more interesting: indeed, there, it turns out that $\tg, \tb, \tp$ are ill-defined. This results in $\tL$ being ill-defined as well. However, given that $\hL$ and $\tL$ only differ by a total derivative, the ill-defined terms of $\tL$ should then be captured by a total derivative only. This is verified in frame A, as can be seen by looking at $\tL$ in \eqref{eq:tLA}. 

Thus for geometric situations, it seems that the NSNS Lagrangian $\hL$ is preferred over the Lagrangian with $Q$-flux $\tL$, since the latter can contain an ill-defined total derivative term. Consequently, the NSNS action is the good action to use, while the integral of $\tL$ is ambiguous. This can seem at odds with the freedom to choose a field basis suggested by GCG. However, from the perspective of string theory this is not surprising, since $\hg, \hB, \hp$ provide the low-energy description of the NSNS sector in geometric settings.

We now turn our attention to the non-geometric configuration of the toroidal example. In frame C, we obtain the famous T-dual of the twisted torus, where the hatted NSNS fields $\hg, \hB, \hp$ are ill-defined and so is the NSNS Lagrangian $\hL$. This agrees with our discussion of non-geometry in ten dimensions. Interestingly, the tilded fields $\tg, \tp$ are well-defined.\footnote{If in one T-duality frame, for instance the A one, the measure $e^{-2\phi} \sqrt{|g|}$ is well-defined, then in any other frame and any basis the dilaton will be well-defined whenever the metric is. Indeed, this measure is invariant under change of frame (action of T-duality, see \eqref{eq:Tddil}) or change of basis (see definition of $\tp$ \eqref{dilmeasure}).} The only tilded field that is ill-defined is $\tb$ (in contrast to $\hB$, we do not allow for gauge transformations of $\tb$, see footnote \ref{foot:gaugetb}). However, since it is $|Q|^2$ and not $\tb$ that appears in $\tL$, this does not result in an ill-defined Lagrangian, see \eqref{eq:tLC}. Similarly to the situation in frame A, the well-defined $\tL$ and ill-defined $\hL$ differ only by a total derivative term as in \eqref{eq:equa}; the NSNS Lagrangian $\hL$ in the non-geometric configuration of frame C can then be written as a well-defined piece plus an ill-defined total derivative term \eqref{eq:hLC}. To summarize, the ill-defined properties of $\hL$ are captured by the total derivative term, together with $\tb$, leaving the $Q$-flux Lagrangian $\tL$ well-defined. Although these features are obtained here in a particular toroidal situation, we think that they are more general and could be observed in other examples as well.\\

Let us remark that in all three T-duality frames of the toroidal example, there is a basis where one field is only locally defined, while the others are well-defined (as discussed further in  appendix \ref{ap:toyex}), and the whole leads to a well-defined Lagrangian. In the geometric frame A, this is true in the hatted basis, where the $B$-field is only defined up to gauge transformations. In frame B, both bases share this property, and the metric connection is the locally defined quantity. In frame C, the same is verified in the tilded basis, where the locally defined field is now $\tb$. Combined with the fact that the Lagrangian $\tL$ is well-defined, this is quite suggestive. We argued already that, for the geometric frames A and B, the hatted NSNS fields form the preferred field basis, since they are well-defined (up to gauge transformations) and lead to a well-defined Lagrangian. Applying the same reasoning to the non-geometric frame C suggests that the tilded field basis should be preferred there. Consequently, considering the well-defined $\tL$ and the associated action in a non-geometric configuration looks like the natural thing to do. However, the rewriting of the ill-defined NSNS Lagrangian leads in addition to an ill-defined total derivative term. The reader may now ask herself if discarding the latter is allowed. This is a subtle question, but let us give some reasons in favour of considering $\tL$ alone. 

A total derivative of non-single-valued fields sometimes contributes as a topological term to the action. In a non-geometric situation though, we know the fields can be made single-valued if we allow some stringy symmetries as transition functions between different patches of the target space. Allowing for such transformations in the integration would make the total derivative term vanish. Thus, by integrating this way over non-geometric fields, we can get rid of this additional term.\footnote{\label{foot:Der}In section \ref{sec:Der}, we mentioned a further possibility: even if the fields require a stringy symmetry to be globally defined, the integrand of the superpotential in \cite{mpt07} is invariant under this symmetry and therefore well-defined. Since we do not consider the same integrand here, we cannot use this property. We also discussed in section \ref{sec:Der} a drawback of this superpotential: it requires the definition of a $Q$-flux in ten dimensions. Given the rewriting done here, such a definition now exists and would be interesting to use in further studies of the superpotential.} Since doing so cures as well the problem of the NSNS action we started with, the equality with the tilded action definitely holds. Nevertheless, there is an advantage in rewriting $\hL$: by isolating the ill-defined terms in a total derivative, we are left with a well-defined Lagrangian (e.g. $\tL$ in frame C of the example) that can be used further.

In addition, we discussed several times that for non-geometric settings, it is not clear that the NSNS action is a good low energy description of string theory. In the toroidal example, a geometric string configuration with a well-defined NSNS Lagrangian leads, via T-dualities, to a non-geometric configuration with ill-defined NSNS Lagrangian but well-defined $Q$-flux Lagrangian. Since T-duality is a symmetry of toroidally compactified string theory, one can think that having a point-like description in one frame (by an NSNS action) should induce another point-like description in another T-duality frame. Similar arguments can be given for the four-dimensional theory and mirror symmetry, see \cite{mpt07} and references therein. For the toroidal example, the natural candidate for this other description is the $\tL$ Lagrangian. Thus, we would shift focus and say that the proper low-energy description of string theory in non-geometric backgrounds is given by $\tg, \tp, \tb$ and not $\hg, \hp, \hB$. As a consequence, we should use the action obtained from $\tL$ to describe the low-energy dynamics of the fields. It would be nice to derive such a claim on the good low-energy description from string world-sheet studies. In particular, although $\tb$ does not appear in the standard string spectrum, it is a rank two antisymmetric tensor as $\hB$ is, and contains the same number of degrees of freedom. More generally, the change of variables from the hatted to the tilded is one-to-one. A world-sheet analysis could give more legitimacy to these new variables as proper physical fields. One approach for such studies could be along the lines of \cite{h06b} or \cite{bhm07}. Another possiblity could be \cite{h08, h09}, see also references therein.

The toroidal example revealed, for each of the three T-duality frames, a basis of fields which are well-defined except for one locally defined field, and these fields lead to a well-defined Lagrangian and action. More generally, it looks reasonable that given a field configuration, geometric or not, there exists a preferred basis for the generalized vielbein $\eee$, in which the quantities entering it lead to a single-valued Lagrangian, that differs from the NSNS $\hL$ at most by a total derivative.\footnote{The existence of such a basis could be related to the notion of generalized parallelizable backgrounds of \cite{gmpw08}. For such backgrounds, it is argued that the $R$-flux should vanish. The toroidal example could be one of such backgrounds (the assumption on $\tb$ is trivially satisfied there, and so we do not have any $R$-flux). We thank D. Waldram for related discussions.} Then we propose the following prescription: in a given configuration, we use the preferred field basis and the associated action to describe the low-energy aspects of string theory.  Here, this is exemplified by choosing between either the hatted basis for a geometric configuration or the tilded basis in a non-geometric situation. But these are only particular examples of basis obtained from a generically $O(2d)$-rotated generalized vielbein. Other choices for the generalized vielbein might lead to a more involved basis, where the degrees of freedom are distributed in both a $B$- and a $\beta$-field.\footnote{Note that in the appendix of \cite{agmp10}, there is an example on a solvmanifold with both a $B$ and $\beta$-field. Note also that the example mentioned in footnote \ref{foot:R} may correspond to a non-geometric situation described in a geometric way by the preferred basis.} The well-defined action would then have all types of fluxes, including $H$ or even $R$. We come back to these possibilities in the following and in the conclusion.\\

The prescription just proposed promotes as the good low-energy description of string theory the well-defined action obtained in the preferred basis. The associated Lagrangian is single-valued, which implies that we can perform on it the usual point-like or ``geometric'' integration. In the case of non-geometry, we can even talk of this Lagrangian as being a ``geometric'' description of a non-geometric situation. In a sense, the rewriting helped us to solve the global issues of the non-geometric situation: by considering only $\tL$ and not the total derivative term, we eventually obtained a globally well-defined action starting from the NSNS action. This allows us to finally make a dimensional reduction in the usual way, and relate in particular the $Q$-flux we obtained to the corresponding term in the four-dimensional scalar potential.  Before doing so, we derive the equations of motion from the single-valued Lagrangian. If the latter is a good low energy effective description of string theory as prescribed, these ten-dimensional equations should correspond to the annihilation of some $\beta$ functions of a sigma model in a non-geometric situation.

\subsection{Ten-dimensional equations of motion}\label{sec:10Deom}

In this section, we derive the equations of motion from a ten-dimensional action with both geometric and non-geometric fluxes. Additionally, these equations are used to investigate whether compactifications to a four-dimensional de Sitter or Minkowski space-time are possible. This analysis reproduces certain constraints from the four-dimensional non-geometry literature.

As discussed above, we focus on the NSNS sector, and consider the proper low-energy description to be given by a preferred basis where the fields $g, \phi, H, Q$ are globally well-defined. This situation is considered here for more generality, and could a priori occur with a more general form of the generalized vielbein in \eqref{eq:HfromE}. This allows us to have both a $B$- and a $\beta$-field, and we restrict ourselves again to the subcase without $R$-flux. Since this basis is different from the hatted and tilded ones considered so far, we drop these notations. As a particular example of this more general  situation, we could consider a configuration with both $Q$- and $H$-fluxes on a compact part of space-time consisting of a non-geometric fiber $\fff$ (along which $Q$ is non-zero), and a geometric base $\bbb$ with a non-zero $H$-flux.

In the presence of an $H$-flux, one can argue that $Q$ is given by \eqref{eq:QH}. Assuming \eqref{eq:intassume}, together with $\beta$ and $Q$ being tensors, it can be shown that the non-geometric $Q$-flux in curved indices is then given by
\beq
{Q_k}^{mn} = \partial_k \beta^{mn} + H_{kpq} \beta^{pm} \beta^{qn} \ . \label{def:Qcurved}
\eeq
In this more general case, neither $\hL$ nor $\tL$ provide a good low-energy description of the string configuration, and we instead propose to use a Lagrangian containing both an $H$- and a $Q$-flux. We then consider the following ten-dimensional action in string frame
\beq \label{eq:10dactionstring}
S = \frac{1}{2 \kappa^2} \int \d^{10} x\ e^{-2\phi} \sqrt{|g|} \left( \R + 4|\d \phi|^2 - \frac{1}{2} |Q|^2 - \frac{1}{2} |H|^2 \right) \ ,
\eeq
where $2\kappa^2= (2\pi)^7 (\alpha')^4$, $\alpha'=l_s^2$ and further conventions are given in appendix \ref{ap:conv}. As we will show in section \ref{sec:4D}, this action is further motivated by reducing to the expected four-dimensional effective field theory. We derive the equations of motion by varying \eqref{eq:10dactionstring} with respect to $g, \phi, B, \beta$ considered as the fundamental fields,\footnote{We take $\beta$ to be independent of $B$ for the derivation of the equations of motion. We believe that in this more general situation, enough restrictions would arise from the Bianchi identities (see section \ref{sec:bi}) to constrain the fields, so that eventually no new degree of freedom is introduced. This reasoning is similar in spirit to the democratic formalism of type II supergravities, where the equations of motion have to be supplied with duality constraints to restore the correct number of degrees of freedom.} and obtain respectively
\bea
0 &= \R_{mn} - \frac{1}{2} g_{mn} \R + 2 g_{mn} \left( |\d\phi|^2 -\nabla^2 \phi \right) + 2 \nabla_m \nabla_n \phi + \frac{1}{4} g_{mn} \left( |H|^2 + |Q|^2 \right) \label{eq:Einstein}\\
&\quad - \frac{1}{4} H_{mpq} {H_n}^{pq} - \frac{1}{4} Q_{mpq} {Q_n}^{pq} + \frac{1}{2} {Q^{p}}_{mq} {Q_{pn}}^{q} \nn \\
0 &= \del_k \left( 8 e^{-2\phi} \sqrt{|g|} g^{km} \del_m \phi \right) 
+ e^{-2\phi} \sqrt{|g|} \left( 2 \R + 8 |\d\phi|^2 - |Q|^2 - |H|^2 \right) \label{eq:eomphi}\\
0 &= \del_k \left( e^{-2\phi} \sqrt{|g|} \left( H^{kmn} + 3 \beta^{p [m} {Q^k}_{pq} \beta^{n] q} \right) \right) \\
0 &= \del_k \left( e^{-2\phi} \sqrt{|g|} {Q^k}_{mn} \right)  - 2 e^{-2\phi} \sqrt{|g|} g_{qr} g_{sn} {H^k}_{pm} \beta^{pq} {Q_k}^{rs} \ ,
\eea
where indices are raised and lowered with $g$. We also trace the ten-dimensional Einstein equation to obtain the Ricci scalar
\beq \label{eq:10einsteintrace}
\R = - \frac{9}{2} \nabla^2 \phi + 5 |\d\phi|^2 + \frac{1}{4} |H|^2 + \frac{3}{4} |Q|^2 \ .
\eeq
Let us make a few remark on the derivation of these equations. The equations of motion for $g, B, \phi$ are the standard ones, to which one adds contributions by $Q$. In particular, the equation for $B$ deviates from its usual form by an extra term containing $\beta$ and $Q$, since $H$ appears in \eqref{def:Qcurved}. For the same reason, $H$ appears in the equation for $\beta$. Note also that the derivation of the equations of motion for $B$ and $\beta$ requires integrations by parts involving total derivatives on $\beta$. Even in the preferred basis, $\beta$ is only locally defined, which means that these total derivatives should be treated with care (see related discussion in \ref{sec:totder}). In the following, we only need the dilaton equation of motion and the Einstein equation, so we disregard these subtleties.\\

Using the above equations, we can now derive conditions that must be fulfilled for compactifications to four-dimensional de Sitter or Minkowski space-time.  For simplicity, we specify the ten-dimensional space-time to be the unwarped\footnote{In type IIB compactifications, this corresponds to taking a large volume limit.} product of a four-dimensional, maximally symmetric space-time, and an internal six-dimensional manifold. Hence,
\beq
g_{mn}(x^m) = \begin{pmatrix} g_{\mu\nu}(x^\mu) & 0 \\ 0 & g_{ij}(x^i) \end{pmatrix} \ ,\label{def:10dmetric}
\eeq
where the ten-dimensional coordinates are denoted $x^{m=0 \dots 9}=(x^{\mu=0 \dots 3},\ x^{i=4 \dots 9})$. Note that being in the preferred basis provides us with a well-defined metric, and a geometric description of the internal space, see section \ref{sec:totder}. We can then speak of an internal manifold. We furthermore choose purely internal fluxes. Using \eqref{eq:Einstein} and \eqref{eq:10einsteintrace}, we then obtain the four-dimensional Ricci scalar
\beq \label{eq:einsteintrace}
\R_4 = g^{\mu\nu} \R_{\mu\nu} 
=2|\d\phi|^2 -\nabla^2 \phi -2 \nabla_{\mu} \nabla^{\mu} \phi - \frac{1}{2} |H|^2 + \frac{1}{2} |Q|^2  \ .
\eeq
Finally, taking the dilaton to be constant,\footnote{In type II supersymmetric solutions, a constant warp factor leads to a constant dilaton. This is not the case in heterotic string.} the requirement of a non-negative four-dimensional curvature then translates into
\beq 0 \leq  2 \R_4 = |Q|^2 -  |H|^2 \ . \label{eq:dSconstraint1} \eeq
Note that the non-geometric $Q$-flux contributes positively to the four-dimensional cosmological constant. For the unwarped metric \eqref{def:10dmetric}, the ten-dimensional Ricci scalar is simply the sum $\R = \R_4 + \R_6$. Combining this with \eqref{eq:10einsteintrace} and \eqref{eq:einsteintrace}, $|H|^2$ can be eliminated in favour of $\R_6$, and $\R_4 \geq 0$ translates into
\beq 0 \leq |Q|^2 - \R_6 \ . \label{eq:dSconstraint2} \eeq

In summary, we find two constraints for the compactifications to a de Sitter or Minkowski vacuum, which can be compared with requirements obtained from a four-dimensional analysis. In section \ref{sec:4D}, we will perform a dimensional reduction allowing us to define a four-dimensional potential \eqref{eq:pot}, which in the vacuum consists of the terms 
\beq
V_{\omega} = -\R_6 \ , \ V_{\bar{H}_3}= \tfrac{1}{2}|H|^2 \ , \  V_Q = \tfrac{1}{2}|Q|^2 \ .
\eeq
Given these definitions, the conditions for a non-negative four-dimensional cosmological constant \eqref{eq:dSconstraint1} and \eqref{eq:dSconstraint2} reproduce equation (5.3) in \cite{cgm09b}. The other terms in this equation should also be reproduced with a similar ten-dimensional reasoning, by including other contributions not considered so far, namely RR fluxes, sources and warping, or the $R$-flux. See for instance \cite{agmp10} for a similar derivation with other degrees of freedom.

The absence of other ingredients has a further consequence in our ten-dimensional analysis. Still assuming a constant dilaton, we can solve the dilaton equation of motion \eqref{eq:eomphi} for $Q$. This solution saturates the inequalities \eqref{eq:dSconstraint1} and \eqref{eq:dSconstraint2}, i.e.
\beq \R_4 = 0 \ , \quad \R_6 = |H|^2 = |Q|^2 \ , \eeq
and so the curvature of the internal manifold is positive and the four-dimensional cosmological constant is zero. Hereby we see that the non-geometric $Q$-flux is just enough to balance the positive curvature of the internal manifold, creating a four-dimensional Minkowski solution. Unfortunately, de Sitter solutions could only be obtained via the inclusion of other ingredients.\footnote{We have checked that the inclusion of an $R$-flux, according to the conjectured action \eqref{eq:QRaction}, does not change this result: we still obtain a four-dimensional Minkowski space-time.}

\subsection{Bianchi identities and the generalized covariant derivative}
\label{sec:bi}

After having obtained the ten-dimensional equations of motion from the rewritten action, we now make a slight digression to discuss how the Bianchi identity $\d \hH = 0$ translates to the tilded basis. In order to do so, we recall from section \ref{sec:Der} that the Bianchi identities of a geometric setting can be recovered as a nilpotency condition on the twisted exterior derivative $\d - H \wedge$. Similarly, it was argued that the Bianchi identities of a non-geometric configuration are obtained as a nilpotency condition for a generalized covariant derivative $\ddd$ that also contains the non-geometric $Q$- and $R$-fluxes. Concretely, this generalized operator acts on differential forms as \cite{stw06}
\beq
\ddd = H\wedge +f\cdot+Q\cdot+R  \llcorner \ ,
\label{eq:gencovder}
\eeq
where $\wedge$ is the standard wedge product,  $\llcorner$ is a contraction  and $\cdot$ is a combination of both. Note that there is no derivative in the definition of $\ddd$ (see for example footnote 3 of \cite{stw06} for a discussion on this point). In particular, in the hatted and tilded field bases of section \ref{sec:computation}, we obtain the operators 
\bea
\label{eq:dhatdtilde}
\hat{\ddd} &= 
\frac{1}{2!} \hat{f}^{a}{}_{bc} \ \he^b \wedge \he^c \wedge \iota_a 
\ +\ \frac{1}{3!} \hH_{abc} \ \he^a \wedge \he^b \wedge \he^c \wedge \ ,
\\
\tilde{\ddd} &=  
\frac{1}{2!} \tilde{f}^{a}{}_{bc}  \ \te^b \wedge \te^c \wedge \iota_a
\ +\ \frac{1}{2!} Q_c{}^{ab} \  \te^c \wedge \iota_a \ \iota_b  \ ,
\nn
\eea
where $\he^a = \he^a{}_m \d x^m$ and similarly for $\te^a$. The contraction is defined by its action on a form as $\iota_a e^b = \delta^b_a$. Note that here and in the following, the flat indices of hatted (tilded) quantities correspond to the hatted (tilded) vielbeine. Since for a given local configuration, the hatted and tilded field bases provide two different descriptions of the same setting (see section \ref{sec:final}), there should be a relation between $\hat{\ddd}$ and $\tilde{\ddd}$. In the rest of this section, we investigate what this relation is.

As in section \ref{sec:computation}, the starting point of our study is the relations \eqref{ghat} and \eqref{Bhat}. Here, we are interested in relations between $\hat{f}^{a}{}_{bc}, \hH_{abc}$ and $\tilde{f}^{a}{}_{bc}, Q_c{}^{ab}$. Assuming that $\hg$ and $\tg$ have the same signature, we deduce from \eqref{ghat} a relation between hatted and tilded vielbeine
\beq
\label{eq:ehattilde}
\he = \te F^{-1} \ ,
\eeq
where we introduce
\beq
F = \id + \tb \tg \ ; \ F^a_{\ b} = \delta^a_{\ b} + \tb^{ac} \eta_{cb} \ ,
\eeq
and we denote by $\eta_{ab}$, in this section only, the tangent space metric (in practice, it would either be the Minkowski metric or the identity; it should not be confused with the GCG metric $\eta$, which does not appear here). Strictly speaking, the relation between the vielbeine is only defined up to an $O(d)$ transformation, which we disregard here. Furthermore, combining \eqref{Bhat} with \eqref{eq:ehattilde} we obtain
\beq
\hB_{ab} = \eta_{ac} \tb^{cd} \eta_{db} \ .
\eeq

From the above equations, in combination with \eqref{struct} and  \eqref{eq:hhat}, it is straightforward to express $\hH_{abc}$ and $\hat{f}^a_{\ bc}$ in terms of tilded quantities:
\bea
\frac{1}{3} \hH_{abc} &= \partial_{[a} \tb_{bc]} - \hat{f}^d_{\ [ab} \tb^{\ }_{c]d} \ , \\
\hat{f}^a_{\ bc} &= \tf^a_{\ bc} + (F^{-1})^a_{\ h} \left( 2 Q_{[b}^{\ \ hd} \eta_{c]d} - \tb^{hd} \eta_{de} \tf^e_{\ bc} - 2 \tb^{de} \eta_{e[c} \tf^h_{\ b]d} \right) \ . \nn
\eea
After a few manipulations, these equalities can be shown to imply 
\beq
\label{eq:fluxequiv}
\eta^{\ }_{d[a} {\hat{f}^d}_{\ bc]} + \frac{1}{3} \hH_{abc} = \eta^{\ }_{d[a} {\tf^d}_{\ bc]} + \eta^{\ }_{d[a} Q^{\ }_{b}{}^{de}\eta^{\ }_{c]e} \ .
\eeq

This equality is very suggestive. Comparing with \eqref{eq:dhatdtilde}, we see that the left-hand side is very similar to $\hat{\ddd}$ and the right-hand side is very similar to $\tilde{\ddd}$. In particular, the contractions $\iota_a$ in \eqref{eq:dhatdtilde} are represented by tangent space metrics $\eta_{ab}$ in \eqref{eq:fluxequiv}. From this similarity we infer that when going from the hatted to the tilded basis, we should replace $\hat{\ddd}$ by $\tilde{\ddd}$.  Thus the NSNS Bianchi identity $\d \hH = 0$, which corresponds to demanding that $\hat{\ddd}^2 = 0$, should translate into the nilpotency condition $\tilde{\ddd}^2 = 0$ in the tilded basis, just as expected from previous studies of Bianchi identities in non-geometric situations \cite{stw06, irw07}, see section \ref{sec:Der}. To put this result on firmer ground it seems necessary to include also the RR sector of the theory, since this sector contains the differential forms on which the generalized exterior derivatives $\hat{\ddd}$ and $\tilde{\ddd}$ operate. We hope to make this more precise in a future publication.

\subsection{Four-dimensional effective field theory}\label{sec:4D}

As discussed in section \ref{sec:bckgd}, string compactifications with non-geometric fluxes have been extensively studied from a four-dimensional perspective. Here, we show that a dimensional reduction of our ten-dimensional action \eqref{eq:10dactionstring} reproduces the expected non-geometric flux terms in the four-dimensional effective theory (in particular its scalar potential). As in section \ref{sec:10Deom}, we assume to be in a preferred basis, where the fields $g, \phi, H, Q$ are globally well-defined, making the Lagrangian single-valued (see section \ref{sec:totder}). This provides us with a geometric (point-like) description of non-geometry, and we can in particular integrate over the internal manifold without problem, and so perform the dimensional reduction.

In the following, we take as a compactification ansatz a metric of the form \eqref{def:10dmetric}. In addition, we take all the fields to depend only on four-dimensional coordinates, and we take the fluxes to be purely internal. The latter are in addition restricted to their vacuum expectation value (denoted in the following with an index $\!\!\!\!\!\phantom{H}^{(0)}$), since we will not consider their fluctuations. We only focus on two moduli,\footnote{In the literature on string compactifications, it is by now customary to also call some scalar fields that have a potential ``moduli''. We stick to this admittedly confusing nomenclature.} namely the volume modulus $\rho$ and the four-dimensional dilaton $\sigma$. These fields are defined by perturbations around vacuum expectation values of the internal volume and the dilaton. In particular, the former is defined as
\beq
g_{ij} \; \rightarrow \; g^{(0)}_{ij} \rho \ ,
\eeq
where the arrow indicates the fluctuation we consider, while the latter is defined as
\beq
e^{-\phi} \; \rightarrow \; e^{-\phi^{(0)}} e^{-\varphi} = g_s^{-1} e^{-\varphi} \ .
\eeq
From the fluctuation $\varphi$ we define the four-dimensional dilaton
\beq 
\sigma = \rho^{3/2} e^{-\varphi} \ .\label{def:4ddilaton}
\eeq
Before we continue, let us emphasize that from the usual ten-dimensional point of view on non-geometry, the volume of a non-geometric space, and hence its volume modulus, are rather ill-defined notions. Some references used this argument to exclude this modulus from the start, or to argue against a large volume limit \cite{hmw02, fww04, stw06, bbvw07}. In this paper, we are not talking of the same object: we performed a change of variables from an ill-defined to a well-defined metric, and the volume modulus is the fluctuation with respect to this new metric.
 
Having defined the moduli, we determine how the terms in \eqref{eq:10dactionstring} scale with respect to them. Taking into account the dependencies on the metric of the various terms, we obtain
\beq
\R_6 \;\rightarrow\; \rho^{-1} \R_6^{(0)}, \quad |H|^2 \;\rightarrow\; \rho^{-3} |H^{(0)}|^2, \quad |Q|^2 \;\rightarrow\; \rho |Q^{(0)}|^2 \ .
\eeq
Additionally, after setting the vacuum expectation value of the internal volume to $L_0^6$, we have
\beq
\int \d^6 x \sqrt{|g_{ij}|} = L_0^6\ \rho^3 \ , \ {\rm where}\ \int \d^6 x \sqrt{|g_{ij}^{(0)}|} = L_0^6 \ .
\eeq

It is now only a matter of putting these definitions together to get the reduced four-dimensional action. For convenience, we drop all indices $\!\!\!\!\!\phantom{H}^{(0)}$ in the following. Also, as we are only interested in the internal curvature and flux contributions, we do not give explicit expressions for the scalar kinetic terms, and denote them as ``kin''. The reduced action is then
\beq
S = \frac{L_0^6}{2 \kappa^2 g_s^2} \int \d^4 x \sqrt{|g_{\mu\nu}|} \rho^3 e^{-2 \varphi} \left( \R_4 + \text{kin} + \rho^{-1} \R_6 - \frac{1}{2} \rho |Q|^2 - \frac{1}{2} \rho^{-3} |H|^2 \right) \ .
\eeq
The four-dimensional Einstein frame action can be obtained by a simple Weyl rescaling with the four-dimensional dilaton \eqref{def:4ddilaton}
\beq
g_{\mu\nu} = \sigma^{-2} g_{\mu\nu}^E \ .
\eeq
Eventually, we find the four-dimensional action in Einstein frame:
\beq
S_E = M_4^2 \int \d^4 x \sqrt{|g_{\mu\nu}^E|} \left( \R_4^E + \text{kin} + \sigma^{-2} \rho^{-1} \R_6 - \frac{1}{2} \sigma^{-2} \rho |Q|^2 - \frac{1}{2} \sigma^{-2} \rho^{-3} |H|^2 \right) \ ,
\eeq
where the four-dimensional Planck mass $M_4$ is given by $M_4^2=L_0^6/(2 \kappa^2 g_s^2)$. The four-dimensional scalar potential can then be read from this action
\beq
\frac{1}{M_4^2}V = -\rho^{-1} \sigma^{-2} \R_6 + \frac{1}{2} \rho \sigma^{-2} |Q|^2 + \frac{1}{2} \sigma^{-2} \rho^{-3} |H|^2 \ . \label{eq:pot}
\eeq
This leads us to the following important result: the potential derived from our ten-dimensional considerations agrees with the four-dimensional arguments in \cite{hktt08}, where the authors propose the dependencies\footnote{Including an $R$-flux, as in the conjectured action \eqref{eq:QRaction}, results in a potential term that scales as in \eqref{eq:Vscale}.}
\beq
V_\omega \sim \sigma^{-2} \rho^{-1}\ ,\quad  V_H \sim \sigma^{-2} \rho^{-3}\ , \quad V_Q \sim \sigma^{-2} \rho\ , \quad V_R \sim \sigma^{-2} \rho^{3}\ .
\label{eq:Vscale}
\eeq

For completeness, let us mention that extremizing the potential, together with the fact that $\rho=\sigma=1$ in the vacuum, leads to
\bea
\frac{1}{M_4^2}\frac{\partial V}{\partial \rho}\Big|_{\rho=\sigma=1} &=  \R_6 + \frac{1}{2} |Q|^2 - \frac{3}{2} |H|^2 = 0 \label{eq:vrho}\\
\frac{1}{M_4^2}\frac{\partial V}{\partial \sigma}\Big|_{\rho=\sigma=1} &=  2 \R_6 - |Q|^2 - |H|^2 = 0 \ .\label{eq:vdil}
\eea
Considering \eqref{eq:vrho} together with \eqref{eq:pot} allows us to recover the two conditions for a non-negative four-dimensional cosmological constant \eqref{eq:dSconstraint1} and \eqref{eq:dSconstraint2}. As in ten dimensions, considering additionally the dilaton contribution (here \eqref{eq:vdil}) leads to the vacuum values
\beq \R_4 = V\big|_{\rho=\sigma=1} = 0, \quad \R_6 = |H|^2 = |Q|^2 \ .\eeq
Thus, the internal manifold has positive curvature and the external space-time is Minkowski, which agrees with the ten-dimensional results of section \ref{sec:10Deom}. As argued there, extra ingredients are needed to obtain a four-dimensional vacuum with non-zero cosmological constant.

\subsection{Double field theory and non-geometry}\label{sec:DFT}

In this section, we make a final digression, and comment on a relation between our rewriting of the NSNS Lagrangian, and a possible double field theory description of non-geometry. Double field theory (DFT) has been introduced in \cite{hz09a, hz09b, hhz10a, hhz10} and developed in a series of subsequent papers. Inspired by earlier results, in particular some work in string field theory, an important achievement of this approach is to obtain a background independent T-duality covariant action for the NSNS sector \cite{hhz10a, hhz10}. Starting from the latter, and given the rewriting performed in this paper, we propose a DFT Lagrangian to describe non-geometric situations.

DFT considers the target space given by doubled geometry (see section \ref{sec:rev}) where one doubles the coordinates, together with the derivatives, so that $O(d,d)$ transformations act linearly on them
\beq
X^M = \begin{pmatrix} y_m \\ x^m \end{pmatrix} \ , \ \del_M =   \begin{pmatrix} \del_y  \\ \del_x \end{pmatrix} \ , \ X' = O X \ , \ \del'=O^{-1} \del \ , \ {\rm for}\ O \in O(d,d) \ ,
\eeq
where as previously we have $m= 1\dots d \ ,\ M= 1\dots 2d$. The NSNS fields (of a $d$-dimensional space-time) are considered in DFT to depend on both sets of coordinates. From these fields, the following action is defined on the doubled space \cite{hhz10}
\bea
  {\cal S}_{DFT} \ & = \ \int \d x \d y \ \lll_{DFT}(\hhh, \Delta) \label{eq:SH}\\
&= \ \int \d x \d y\ e^{-2\Delta} \Big( \frac{1}{8}\ \hhh^{MN}\partial_{M}\hhh^{KL}
  \ \partial_{N}\hhh_{KL}-\frac{1}{2}\hhh^{MN}\partial_{N}\hhh^{KL}\ \partial_{L} \hhh_{MK}\nn\\
  & \qquad\qquad \qquad \qquad -2\ \partial_{M}\Delta\ \partial_{N}\hhh^{MN}+4\hhh^{MN}\,\partial_{M}\Delta\ \partial_{N}\Delta \Big)\ , \nn
\eea
where $e^{-2\Delta}= e^{-2\phi} \sqrt{|g|}$ is the standard NSNS measure, and $\hhh(\hg, \hB)$ is the generalized metric given (up to a change of conventions) by \eqref{genmet}. In addition to an interesting gauge symmetry, this action has the property to be written in a covariant way under an $O(d,d)$ transformation. Indeed, that $\Delta$ is invariant under such a transformation can be seen from \eqref{eq:Tddil}, and the same goes for $\d x \d y=\frac{1}{2} \d X^T \eta \d X$. We also know that $\hhh$ is transforming (bi)linearly as in \eqref{eq:HO}, so each term of the action is clearly invariant.

This action is of particular interest, because if the fields are now considered to depend only on the standard coordinates $x^m$, one recovers the standard NSNS action (see also \cite{ms92}). More precisely, setting $\del_y = 0$, one gets
\beq
\xymatrix{ \lll_{DFT}(\hhh(\hg, \hB), \hat{\Delta})\ \ar@{=}[r]^{\ \ \ \ \ \del_y = 0}&\ \hL + \del(\dots) } \ , \label{eq:DFTdel}
\eeq
where we recall from \eqref{eq:intlag} that $\hL$ is the standard NSNS Lagrangian; the latter is then obtained up to a total derivative term \cite{hhz10}.\\

Combining this result with those of our paper, we deduce the following diagram
\beq
 \xymatrix{ \lll_{DFT}(\hhh(\hg, \hB), \hat{\Delta})\ \ar@{=}[r]^{\eqref{eq:intequality}}_{\eqref{dilmeasure}} & \ \lll_{DFT}(\hhh(\tg, \tb), \tilde{\Delta}) \\
                               \hL+ \del(\dots)\ \ar@{=}[u]^{\eqref{eq:DFTdel}} \ar@{=}[r]^{\eqref{eq:intlag}} &\ \tL+ \del(\dots) \ar@{==}[u] } \label{eq:diagr} 
\eeq
The left column is the result of DFT \cite{hhz10} that we gave in \eqref{eq:DFTdel}. The bottom line is the equality of Lagrangians derived in this paper. The top line is the proposal we make: since our change of variables from the hatted to the tilded basis leaves the generalized metric $\hhh$ and $\Delta$ invariant (see \eqref{eq:intequality}, \eqref{dilmeasure}), one should consider the DFT Lagrangian given in \eqref{eq:SH}, but where $\hhh$ and $\Delta$ are expressed in terms of the tilded variables $\tg, \tb, \tp$.\footnote{Given the discussion made in section \ref{sec:totder}, one should in general consider expressions in terms of the fields in the preferred basis, where the Lagrangian is well-defined. In any case, the formal $\lll_{DFT}$ will still be given by \eqref{eq:SH}, since $\hhh$ and $\Delta$ should remain invariant.} This Lagrangian (in the top-right corner of \eqref{eq:diagr}) should provide an interesting DFT description of non-geometry, up to global issues that we discussed in section \ref{sec:totder}. In addition, the chain of equalities in \eqref{eq:diagr} shows that the Lagrangians in the last column should be equal, up to a total derivative term, once one considers $\del_y = 0$. Therefore, the DFT Lagrangian $\lll_{DFT}(\hhh(\tg, \tb), \tilde{\Delta})$ should give back the non-geometric $Q$-flux in the ten-dimensional Lagrangian, provided the assumption \eqref{eq:intassume} is satisfied. It would be interesting to study this DFT Lagrangian further, for instance its properties under DFT gauge transformations.

\section{Conclusion} \label{sec:conclusion}

Although the idea that string theory compactifications are possible also on non-geometric spaces is not new, the understanding of non-geometry is still incomplete. Over the years, this subject has been studied from different angles. From a four-dimensional effective field theory approach, non-geometry was first described in terms of non-geometric $Q$- and $R$-fluxes, which are needed in order to write the four-dimensional theory covariantly with respect to T-duality. On the other hand, from the ten-dimensional perspective, non-geometry has primarily appeared as the failure of the NSNS fields to be globally well-defined, unless a stringy symmetry is used to patch them. Since these global issues prevent a standard dimensional reduction, no straightforward relation between the four- and ten-dimensional descriptions of non-geometry has been obtained. In particular, no ten-dimensional interpretation of the non-geometric fluxes has been given.

In this paper we make progress in establishing such a relation.  Our work is inspired by Generalized Complex Geometry (GCG), where it has been shown that it is always possible to perform a local change of field variables, replacing the NSNS metric $\hg$ and $B$-field $\hB$ by a  new metric $\tg$ and an antisymmetric bivector $\tb$. This change of field basis is permitted by an $O(2d)$ action that leaves the generalized metric (a central object in GCG) invariant \eqref{eq:intequality}. Defining as well a new dilaton $\tp$ to replace the NSNS dilaton $\hp$, we then rewrite the ten-dimensional NSNS Lagrangian $\hL$ in terms of the tilded fields. Particularly, by using a simplifying assumption \eqref{eq:intassume} we show that 
\beq
\nn
\hL= e^{-2\hp} \sqrt{|\hg|} \left(\hR + 4|\d \hp|^2 - \frac{1}{2} |\hH|^2 \right)= e^{-2\tp} \sqrt{|\tg|} \left(\tR + 4|\d \tp|^2 - \frac{1}{2} |Q|^2 \right) + \del (\dots) = \tL + \del (\dots) \ ,
\eeq
where $\del (\dots)$ is a total derivative term, and the terms that depend on $\tb$ in $\tL$ are collected in $|Q|^2$. Concretely, the rewriting results in a $Q$ that is related to $\tb$ as in \eqref{eq:nongeofluxsimple}. Interestingly, this agrees with relations \eqref{nongeoflux}, \eqref{eq:QH}, that have been derived in the literature from algebraic considerations \cite{gmpw08, h09}. It should be emphasized that  this reformulation of the theory is obtained without introducing any new degree of freedom; indeed we have only worked out what the change of field variables implies on the level of the Lagrangian. Nevertheless, rewriting the Lagrangian results in a ten-dimensional formulation of the theory that contains a term which could correspond to the four-dimensional  non-geometric $Q$-flux.

The rewriting of the Lagrangian is valid in both geometric and non-geometric settings. However, the most interesting aspects of the procedure become apparent when considering non-geometric configurations, where the NSNS fields and Lagrangian $\hL$ are typically ill-defined.  For such configurations, the tilded metric, dilaton and $Q$-flux may very well be single-valued, leading to a well-defined Lagrangian $\tL$. Indeed, this is the case for a well-known non-geometric toroidal configuration. Consequently,  the action $\tilde{S}$ associated with $\tL$ is well-defined, and can be straightforwardly dimensionally reduced. After a discussion on global aspects, we perform this exercise, and finally obtain a link between the ten- and four-dimensional perspectives of non-geometry, at least for the cases  that fulfill \eqref{eq:intassume}.  In particular, we show that the obtained ten-dimensional $Q$-flux does reduce to its four-dimensional counterpart. 

Furthermore, we advocate that a similar well-defined field basis and Lagrangian should exist for any configuration, geometric or not. We then prescribe that this preferred field basis should be used for the low-energy description of string theory. An argument in favour of this  prescription is that if a  geometric string configuration has a low-energy Lagrangian description, then such a point-like description should also exist for any dual of the configuration. In the toroidal example, we find that the different preferred field bases in geometric and non-geometric T-dual frames, lead to equal and well-defined Lagrangians. It would be interesting to find a string world-sheet derivation of these arguments. For this purpose, we derive the ten-dimensional equations of motion that should correspond to the vanishing of the $\beta$ functions of a world-sheet CFT.

As side results, we apply the change of field basis to the NSNS Bianchi identity. This constraint on the $H$-flux can be reformulated as the nilpotency condition of a twisted derivative. Similarly, it has been argued that demanding the nilpotency of a generalization of this derivative should reproduce the flux constraints in non-geometric settings. Through our rewriting, we find support for this idea. We also discuss a possible double field theory description of non-geometry, and propose a Lagrangian which could serve this purpose. It would be interesting to further pursue these studies.\\

In this paper we have restricted our studies of the NSNS action to the simplified setting where $\tb$ fulfills \eqref{eq:intassume}. In particular, this means that we have been focusing on locally geometric configurations with zero $R$-flux. Rewriting the NSNS action without the simplifying assumption should also be possible, though technically more involved, and we conjecture that the following holds:
\beq
\label{eq:QRaction}
\hL= e^{-2\hp} \sqrt{|\hg|} \left(\hR + 4|\d \hp|^2 - \frac{1}{2} |\hH|^2 \right)=e^{-2\tp} \sqrt{|\tg|} \left(\tR + 4|\d \tp|^2 - \frac{1}{2} |Q|^2 - \frac{1}{2} |R|^2 \right) + \partial(...)\ , 
\eeq
where the non-geometric fluxes should correspond to those defined in \eqref{nongeoflux}. Proving this conjecture is work in progress \cite{a11b}. Again, it is imaginable that the rewritten Lagrangian on the right-hand side of \eqref{eq:QRaction} is well-defined for some configuration, and can then be used for the dimensional reduction of the theory. It is also possible that both a $B$- and a $\beta$-field are present simultaneously in a preferred basis, adding an $H$-flux term to the above Lagrangian. In that case, there should be extra constraints to restore the correct number of degrees of freedom.

It would be interesting to extend our work to study the effects of non-geometry on the RR sector, for example along the lines of \cite{ms07}.
Furthermore, some four-dimensional arguments suggest that other non-geometric fluxes are needed in the RR sector, in addition to the NSNS $Q$- and $R$-fluxes. Indeed, the introduction of $Q$- and $R$-fluxes spoils the $S$-duality invariance of type IIB supergravity, and in order to reinstate it, a new set of fluxes should be included  \cite{acfi06, acr09, dlr10}. Such fluxes are not contained in GCG, which focuses on the NSNS sector, but they could be described in a framework such as Exceptional Generalized Geometry (EGG) \cite{h07, pw08, glsw09, aacg10, go11}, or following \cite{bp10, bgp11}. It would be interesting to investigate whether there is a field redefinition  that would also make the non-geometric fluxes of the RR sector appear in the ten-dimensional action, in parallel to the discussion we have made here. 

One of the main advantages of reformulating the NSNS sector of ten-dimensional supergravity in terms of a new basis of fields is that,  also for non-geometric configurations, it can provide a globally well-defined Lagrangian which integrates to a well-defined action. This action can be used to derive ten-dimensional equations of motion for the fields, and it can also be dimensionally reduced to four dimensions. This is promising for the construction of phenomenologically interesting  four-dimensional theories and solutions, which is one of the main reasons for studying non-geometry. Relating our rewriting to the GCG tools, in particular the $\beta$-transform on pure spinors, should help in this respect \cite{mpz06,mpt07}. 

Also, in four-dimensional gauged supergravities, it has been shown that non-geometric fluxes contribute positively to the four-dimensional cosmological constant, and that de Sitter solutions can then be found \cite{cgm09a,cgm09b}. Through our dimensional reduction, we find that the $Q$-flux gives the expected positive contribution to the cosmological constant. Although this is encouraging for the construction of de Sitter solutions, a detailed inspection shows that various fluxes balance and the four-dimensional space-time is Minkowski. Thus, while focusing on the NSNS sector is enough to make the non-geometric fluxes appear in the ten-dimensional formulation of the theory, the lack of additional ingredients prevents us from making contact with concrete four-dimensional de Sitter solutions. 

Finally, it would be interesting to relate $\tL$ to the non-commutative and non-associate geometry, discussed in 
\cite{Blumenhagen:2010hj,Lust:2010iy,Blumenhagen:2011ph}, where it was conjectured that the effective action of a non-geometric closed string background is described by a non-associative version of gravity. We hope to come back to these questions in future publications.

\section*{Acknowledgements}

It is a pleasure to thank N. Carqueville, R. Minasian and D. Waldram for interesting discussions. The research of M.~L. and D.~L.  was supported by the Munich Excellence Cluster for Fundamental Physics ``Origin and the Structure of the Universe''.


\newpage
\begin{appendix}

\section{Conventions}\label{ap:conv}

In this appendix we summarize the conventions and notations we use.

\subsection*{Metric and structure constants}
A metric $g$ of a $d$-dimensional space-time is expressed in a local basis of one-forms $\{\d x^{m=0 \dots d-1} \}$ as
\beq
\d s^2= g_{mn} \d x^m \d x^n \ .
\eeq
We then denote by $g$ the $d \times d$ matrix of coefficients $g_{mn}$. The absolute value of its determinant is denoted $|g|$. The metric can also be
 expressed in terms of the vielbein matrix $e$, of coefficient $e^a_{\ m}$, as
\beq
g=e^T \id_d\ e \ ,
\eeq
when it is positive definite. A structure constant, with respect to a given vielbein $e^a_{\ m}$ and its inverse $e_a^{\ m}$, is given by 
\beq
f^a_{\ bc}=e^a_{\ m} \left(e_b^{\ k} \del_k e_c^{\ m}- e_c^{\ k} \del_k e_b^{\ m} \right) = - 2 e_{[b}^{\ \ k} e_{c]}^{\ m} \del_k e^a_{\ m} \ .\label{struct}
\eeq

\subsection*{Forms and fluxes}
Our convention for a $p$-form $A$ on a basis of one-forms $\{\d x^{m}\}$ is
\beq
A=\frac{1}{p!}A_{m_1\dots m_p}\d x^{m_1}\w \dots \w \d x^{m_p} \ .
\eeq
In particular, for the two-form $B$-field, we denote by $\hB$ the matrix of coefficients $\hB_{mn}$. A wedged form given by the wedge product of $A$ and a
 $q$-form $B$ is defined as
\beq
\frac{1}{(p+q)!}(A\w B)_{m_1\dots m_{p+q}}=\frac{1}{p!q!}A_{[m_1\dots m_p} B_{m_{p+1}\dots m_{p+q}]} \ , \label{coefwedge}
\eeq
where the right-hand side indices are totally antisymmetrized. The antisymmetrization of two one-forms $a$ and $b$ is given by
 $a_{[m}b_{n]} = \frac{1}{2!} (a_m b_n - a_n b_m)$, and so on for forms of higher degrees. For a given metric $g_{mn}$, we denote
\beq
|A|^2=\frac{A_{m_1 \dots m_p} A^{m_1 \dots m_p}}{p!}=\frac{1}{p!}A_{m_1 \dots m_p} A_{n_1 \dots n_p} g^{m_1 n_1} \dots g^{m_p n_p} \ .\label{eq:A2}
\eeq
This applies in particular to $\d\hp$ and $\hH$ in the NSNS action. For the non-geometric fluxes (taken as tensors), we denote similarly
\beq
|Q|^2= \frac{1}{2!} {Q_k}^{mn} {Q_p}^{qr} g^{kp} g_{mq} g_{nr}  \ , \ |R|^2 = \frac{1}{3!} R^{kmn} R^{pqr} g_{kp} g_{mq} g_{nr} \ . \label{eq:Q2R2}
\eeq

\section{T-dualities on a toroidal example}\label{ap:toyex}

In this section, we illustrate our general results with the well-known toroidal example, where one applies T-dualities on a three-torus with non-zero $B$-field. This simple toy example has been discussed at length in the literature (see section \ref{sec:bckgd}), and serves here as a pedagogical introduction to the different bases for the generalized vielbeine and associated fields, in a given T-duality frame. In particular, we write down the explicit field configurations, both in the $\hB$- and  $\tb$-basis (the hatted and tilded basis), for three different T-duality frames.\footnote{Although these configurations do not solve the supergravity equations of motion, one can view the three-dimensional fields and Lagrangian as being components of ten-dimensional ones. Their ten-dimensional completions can then be solutions.} We then plug these fields into the hatted and tilded Lagrangians of equation \eqref{eq:intlag} and check if they are well-defined. We comment on the results in a final summary.

Our starting point is a square three-torus with coordinates $x, y, z$, that are periodically identified 
\beq
(x, y, z) \sim (x + 2 \pi, y, z) \sim (x, y + 2 \pi, z) \sim (x, y, z + 2 \pi) \ .
\eeq
We assume that no field depends on $x$ and $y$, so T-dualizing along these directions is allowed according to the Buscher rules \cite{b87, b88}.  It is well-known that combining these two transformations leads to a configuration that is not globally geometric \cite{kstt02, lnr03, stw05}. As we discuss in more detail below (see also \eqref{Tdchain}),  we have the following T-duality chain:
\bea
A: H_{xyz} \ \ 
\buildrel {x} \over \longleftrightarrow \ \ 
B: f^{x}{}_{yz} \ \ 
\buildrel {y} \over \longleftrightarrow \ \ 
C: {Q_z}^{xy} \ \ 
\buildrel {z} \over \longleftrightarrow \ \ 
D: R^{xyz} \ ,\nn
\eea
where $A, B, C, D$ denote the T-duality frames and $H, f, Q, R$ are background fluxes. The non-geometric configurations occur in frames C and D; the former is still locally geometric, whereas the latter lacks even a local geometric description \cite{stw05, dh05}. The last T-duality, along $z$, is strictly speaking not allowed by the Buscher rules, since there is no isometry in this direction. However, it can be argued that T-dualizing along $z$ still makes sense (see footnote \ref{foot:TdR}), and leads to a configuration with an $R$-flux. In the rest of this section, we focus on the first three T-duality frames, and describe their properties in detail. Our observations are summarized in Table \ref{tab:Tframes} at the end of the section.

\subsection*{Frame A}
We start with a square three-torus, with metric $(\hg_A)_{ij} = \delta_{ij}$ along $(x,y,z)$ directions, and a dilaton and $B$-field with non-trivial $z$ dependence:
\beq
\hB_A = \begin{pmatrix}  0 & -z & 0 \\ z & 0 & 0 \\ 0 & 0 & 0 \end{pmatrix}  
\ ,
\quad
\hp_A = \phi(z)
\ .
\eeq
Here, the index $A$ on the fields denotes the name of the T-duality frame, and the fields are all in the hatted basis. All fields are globally well-defined when gauge transformations of $\hB$ are allowed, so this is a geometric configuration. Consequently, the terms in the $\hB$-basis Lagrangian are all well-defined: 
\beq \label{eq:hLA}
\hat{\cal L}_A= 
e^{-2\hp_A} \sqrt{|\hg_A|} \left(\hR_A + 4|\d \hp_A|^2 - \frac{1}{2} |\hH_A|^2 \right)= 
e^{-2\phi}  \left(-\frac{1}{2} + 4|\d \phi|^2  \right) \ .
\eeq

Switching to the $\tb$-basis (using the formulas of section \ref{sec:def}) the fields are
\beq
\tg_A = \begin{pmatrix} 1+z^2 & 0 & 0 \\ 0 & 1+z^2 & 0 \\ 0 & 0 & 1 \end{pmatrix}  
\ ,
\quad
\tb_A =\frac{1}{1+z^2} \begin{pmatrix}  0 & -z & 0 \\ z & 0 & 0 \\ 0 & 0 & 0 \end{pmatrix}  
\ ,
\quad
\tp_A = \phi(z) + \frac{1}{2} \ln (1 + z^2)
\ .
\eeq
Neither of these fields are well-defined, as they do not respect the periodicity of the $z$ coordinate. Using \eqref{eq:asnongeoflux} and \eqref{eq:Q2R2}, the terms in the $\tb$-basis Lagrangian can be computed:
\beq \label{eq:tLA}
\tilde{\cal L}_A= 
e^{-2\tp_A} \sqrt{|\tg_A|} \left(\tR_A + 4|\d \tp_A|^2 - \frac{1}{2} |\tQ_A|^2 \right)= 
e^{-2\phi}  \left(-\frac{1}{2} + 4|\d \phi|^2  \right) + 
4 \partial_{z}  \left(e^{-2\phi} \tb^{xy} \right)\ ,
\eeq
which differs from the Lagrangian in the $\hB$-basis by a total derivative.  Note that the total derivative term is ill-defined, since $\tb_A$ does not respect the periodicity of $z$. In fact, integrating this term over $z$ gives
\beq \label{eq:boundaryterms}
\int_0^{2\pi} d z \partial_{z}   \left(e^{-2\phi} \tb^{xy} \right)
= -\frac{2 \pi}{1+4 \pi^2} e^{-2\phi (2 \pi)} 
\ ,
\eeq
which is clearly non-vanishing.  The total derivative term can also be computed directly from the general formula \eqref{eq:totalderivative}, with the same result.

\subsection*{Frame B}
T-dualizing along $x$ leads to a twisted torus with zero $B$-field and unchanged dilaton: 
\beq
\hg_B = \begin{pmatrix} 1 & z & 0 \\ z & 1+z^2 & 0 \\ 0 & 0 & 1 \end{pmatrix}   
\ ,
\quad
\hB_B = 0 
\ ,
\quad
\hp_B = \phi(z) 
\ .
\eeq
This metric is globally well-defined, provided one allows for gauge transformations on the off-diagonal connection. The configuration is then geometric, as we can also see through the volume of the twisted torus being single-valued ($\mbox{det} \hg_B = 1$). Consequently, the terms in the Lagrangian are all well-defined:
\beq
\hat{\cal L}_B= 
e^{-2\hp_B} \sqrt{|\hg_B|} \left(\hR_B + 4|\d \hp_B|^2 - \frac{1}{2} |\hH_B|^2 \right)= 
e^{-2\phi}  \left(-\frac{1}{2} + 4|\d \phi|^2  \right) \ .
\eeq

Since the $B$-field is trivial, so is the $\beta$-field, and there is no difference between the hatted and tilded fields and  Lagrangians:
\beq
\tg_B = \hg_B  
\ ,
\quad
\tb_B = \hB_B = 0 
\ ,
\quad
\tp_B = \hp_B = \phi(z) 
\ ,
\quad
\tilde{\cal L}_B = \hat{\cal L}_B
\ .
\eeq
In particular, both the hatted and tilded Lagrangians are well-defined.

\subsection*{Frame C}
We now perform a second T-duality, this time along $y$. In the $\hB$-basis, we have the following fields: 
\beq \label{eq:hfieldsC}
\hg_C =  \frac{1}{1+z^2} \begin{pmatrix} 1 & 0 & 0 \\ 0 & 1 & 0 \\ 0 & 0 & 1+z^2 \end{pmatrix} 
\ ,
\quad
\hB_C = \frac{1}{1+z^2}  \begin{pmatrix}  0 & z & 0 \\ -z & 0 & 0 \\ 0 & 0 & 0 \end{pmatrix}
\ ,
\quad
\hp_C = \phi(z)  - \frac{1}{2} \ln (1+z^2)
\ .
\eeq
All these fields are globally ill-defined, as is the determinant of the metric, $\mbox{det} \hg_C = \frac{1}{(1+z^2)^2}$. But they can be patched, when going around the base circle, by a T-duality transformation which makes them single-valued \cite{lnr03}. Thus, this is a non-geometric configuration. Computing the terms in the Lagrangian, we obtain
\beq \label{eq:hLC}
\hat{\cal L}_C= 
e^{-2\hp_C} \sqrt{|\hg_C|} \left(\hR + 4|\d \hp|^2 - \frac{1}{2} |\hH|^2 \right)= 
e^{-2\phi}  \left(-\frac{1}{2} + 4|\d \phi |^2 \right) + 
4 \partial_{z}  \left(e^{-2\phi} \hB_{xy} \right) \ .
\eeq
Note that the total derivative term is not single-valued, and in fact integrates to the right-hand side of \eqref{eq:boundaryterms}, up to a sign. This term can also be computed using the general expression for the total derivative term \eqref{eq:totalderivative}.

If we instead use the tilded basis, this configuration comprises the fields 
\beq \label{eq:tfieldsC}
\tg_C =   \begin{pmatrix} 1 & 0 & 0 \\ 0 & 1 & 0 \\ 0 & 0 & 1 \end{pmatrix}
\ ,
\quad
\tb_C =  \begin{pmatrix}  0 & z & 0 \\ -z & 0 & 0 \\ 0 & 0 & 0 \end{pmatrix} 
\ ,
\quad
\tp_C = \phi(z) 
\ .
\eeq
Here $\tg_C$ and $\tp_C$ are both well-defined, so the non-geometricity only shows up in the non-trivial $\tb_C$, whose form was first derived in \cite{gs06}. The non-zero components of $\tQ_C$ are $(\tQ_C)_z{}^{xy} = -(\tQ_C)_{z}{}^{yx} = 1$. Thus, the Lagrangian only contains well-defined terms
\beq \label{eq:tLC}
\tilde{{\cal L}}_C= 
e^{-2 \tp_C} \sqrt{|\tg_C|} 
\left(\tR_C + 4|\d \tp_C|^2 - \frac{1}{2} |\tQ_C|^2 \right) 
= e^{-2 \phi}  
\left( 4|\d \phi|^2 - \frac{1}{2}  \right)
\  .
\eeq

\subsection*{Summary}

\begin{table}[tb]
\begin{tabular}{l l l}
\hline
T-duality frame & Fields and Lagrangian ($\hB$-basis) & Fields and Lagrangian ($\tb$-basis)\\
\hline
\hline
A ($H_{abc}$) & $\hg_A$, $\hp_A$ ({\sc G}), $\hB_A$ ({\sc L})  & $\tg_A$, $\tb_A$, $\tp_A$ ({\sc L})\\
& $\hat{\cal L}_A= {\cal L} $
& $\tilde{\cal L}_A= {\cal L} + 4 \partial_{z}  \left(e^{-2\phi} \tb^{xy} \right)$\\
\hline
B ($f^a{}_{bc}$) & $\hg_B$, $\hB_B = 0$, $\hp_B$ ({\sc G}) & $\tg_B$, $\tb_B = 0$, $\tp_B$ ({\sc G})\\
& $\hat{\cal L}_B= {\cal L} $
& $\tilde{\cal L}_B= {\cal L}$\\
\hline
C (${Q_c}^{ab}$) & $\hg_C$, $\hB_C$, $\hp_C$ ({\sc L}) & $\tg_C$, $\tp_C$ ({\sc G}), $\tb_C$ ({\sc L})\\
& $\hat{\cal L}_C= {\cal L} + 4 \partial_{z}  \left(e^{-2\phi} \hB_{xy} \right)  $
& $\tilde{\cal L}_C= {\cal L}$\\
\hline
\end{tabular}
\caption{T-duality frames and field bases for the three-torus toy example. Here ``({\sc G})'' means that a field is globally well-defined, and ``({\sc L})'' means that a field is only locally defined. The Lagrangian ${\cal L}$ is globally well-defined, whereas the total derivative terms contain locally defined fields. \label{tab:Tframes}}
\end{table}

The configurations in the various T-duality frames and field bases for this toy model are summarized in Table \ref{tab:Tframes}. Let us highlight some interesting aspects of this example.

First, note that for each frame the difference between the hatted and tilded Lagrangians is a total derivative, in agreement with the results of section \ref{sec:computation}. Indeed, we have checked that \eqref{eq:totalderivative} gives the right total derivative terms for this example. However, note also that the total derivative terms are ill-defined, and it is not clear whether they can be neglected; when integrating them we pick up boundary terms \eqref{eq:boundaryterms}. 

Second, note that for the different T-duality frames of this example, there is a preferred field basis in which all terms in the Lagrangian are well-defined. The preferred basis is the $\hB$-basis for a geometric configuration (A), and the $\tb$-basis for a non-geometric configuration (C). For frame B, where $B=\beta=0$, both bases yield well-defined terms in the Lagrangian. Furthermore, when expressed in these preferred field bases, Table \ref{tab:Tframes} shows that the Lagrangians are the same in the three frames (they all equal $\lll$). 

Finally, using this preferred field basis in every T-duality frame, we see that each time, one field is only locally defined, while the others are well-defined. This local quantity, which basically equals $z$, moves from one field to the other while acting with T-duality and changing frames. Indeed, we could write a T-duality chain of the local quantity
\bea
& \hB_{xy} \xrightarrow{x} {A^x}_y \xrightarrow{y} \tb^{xy} \ ,\label{eq:chainlocal}
\eea
where ${A^x}_y \d y$ is the connection one-form which encodes the non-trivial fibration of the twisted torus in frame B: the metric can be written as $\d x +{A^x}_y \d y$. This chain is reminiscent in many ways of the chain \eqref{Tdchain}, where one also lifts the index along the T-duality direction. 

All these observations inspired us to make a particular prescription, that we discuss at length in section \ref{sec:totder}.

\section{Computational details}\label{ap:comp}

In this appendix, we give technical details on the rewriting of the NSNS Lagrangian, discussed in section \ref{sec:computation}. As a convention, a
 derivative only acts on the first object on its right, unless we put some brackets. Let us also mention that we extensively use the following relation, on any
 invertible matrix $A$ of coefficient $A^{mp}$ 
\beq
A^{mp}\left(\partial_k A^{-1}_{pn}\right)=-A^{-1}_{pn}\left(\partial_k A^{mp}\right)\ .
\eeq

\subsection{The Ricci scalar}

For a generic metric $g_{mn}$ with Levi-Civita connection, one has for the connection coefficients
\beq
2\G_{mkn}=\left(\del_k g_{mn}+\del_n g_{mk}-\del_m g_{kn}\right) \ ,\ {\G^p}_{kn}=g^{pm} \G_{mkn}={\G^p}_{nk} \ , \ {\G^{pq}}_{n}=g^{qk} {\G^p}_{kn} \ .
\eeq
Then the Ricci scalar is given by 
\beq
\R=g^{ln} \del_k{\G^k}_{nl} - g^{lm} \del_m {\G^k}_{kl} + {\G^{pn}}_n {\G^k}_{kp} - {\G^{pn}}_k {\G^k}_{np} \ .\label{eq:defR}
\eeq
Computed explicitly in terms of the metric, each of these terms and the Ricci scalar are then given by 
\bea
g^{ln}\del_k{\G^k}_{nl} =& g^{lm}g^{ku} \del_k \del_m g_{lu} - \frac{1}{2} g^{lu}g^{km} \del_k\del_m g_{lu} \nn\\
&- \del_k g_{rs} \del_m g_{lu} \left(g^{lm}g^{kr}g^{su}  - \frac{1}{2} g^{lu}g^{kr}g^{sm}  \right) \ ,\label{t1} \\
g^{lm} \del_m {\G^k}_{kl} =& \frac{1}{2} g^{km} g^{lp} \del_m\del_k g_{lp} - \frac{1}{2} \del_m g_{rs} \del_k g_{lu} g^{km} g^{lr} g^{su} \ ,\label{t2} \\
{\G^{pn}}_n {\G^k}_{kp} =& \frac{1}{4} \del_m g_{nl} \del_k g_{pu} \left( 2 g^{kl} g^{mn} g^{pu} - g^{km} g^{ln} g^{pu} \right) \ ,\label{t3} \\
{\G^{pn}}_k {\G^k}_{np} =& \frac{1}{4} \del_m g_{nl} \del_k g_{pu} \left( 2 g^{kl} g^{pm} g^{nu} - g^{km} g^{lp} g^{nu} \right) \ ,\label{t4} \\
\R=& g^{lm}g^{ku} \del_k \del_m g_{lu} - g^{lu}g^{km} \del_k\del_m g_{lu} \label{tR}\\
& +\frac{1}{2} \del_m g_{ln} \del_k g_{pu} \Big( 2 g^{kl} g^{mn} g^{pu} - \frac{1}{2} g^{km} g^{ln} g^{pu} \nn\\
& \ \ + \frac{3}{2} g^{km} g^{np} g^{lu} - g^{mp} g^{kn} g^{lu} - 2 g^{mn} g^{kp} g^{lu} \Big) \ .\nn
\eea

Using \eqref{tR}, one directly gets the expression \eqref{Rtilde} for the Ricci scalar $\tR$ of $\tg$. Computing $\hR$ in terms of $\tg$ and $\tb$ is more
involved. To do so, we consider each of the four terms of \eqref{eq:defR}. We compute each of them by replacing in it $\hg$ by its expression
 \eqref{eq:convrewr}, and making use of the assumption \eqref{eq:assume} and the simplifications \eqref{eq:simplif}. We put everything together and obtain
\bea
\hR-\tR=& - \partial_{m}\tg_{np} \partial_{k}\tg_{rs} \left(2\tg^{km}\tg^{nr}\tg^{ps}+2\tg^{rs}\tg^{mn}\tg^{pk}+\frac{1}{2}\tg^{ms}\tg^{nr}\tg^{pk}\right) \label{R-Ras0}\\
& -\tg_{ln} \partial_{k}\tb^{kl} \partial_{m}\tb^{mn} -\frac{1}{2}\tg_{ln} \partial_{k}\tb^{lm} \partial_{m}\tb^{nk}  \nn\\
& +2\tg^{km}\tg^{pq} \partial_{k}\partial_{m}\tg_{pq} +2\tg^{km}\iG_{pq} \partial_{k}\partial_{m}G^{qp} \nn\\
& + \partial_{m}G^{np} \Big(-2\tg^{rs}\tg^{km}\iG_{sn} \partial_{k}\tg_{pr} +2\tg^{qs}\tg^{mr}\iG_{sn}\partial_{p}\tg_{qr} - 2 \tg^{mr} \tg^{ks} \iG_{pn} \del_k \tg_{rs} \nn\\
& \quad-\tg^{rs}\tg^{km} \iG_{pn} \partial_{k}\tg_{rs} -\tg^{mr}\tg^{qs}\iG_{ps} \partial_{n}\tg_{qr}  +\tg^{km}\tg^{rs}\iG_{ps} \partial_{k}\tg_{nr} \Big) \nn\\
& + \partial_{m}G^{np} \Big(\iG_{qn} \partial_{p}G^{mq} +\hg_{qn} \partial_{p}G^{qm}  -\frac{1}{2} \hg_{qp} \partial_{n}G^{mq} \Big) \nn\\
& + \partial_{m}G^{np}  \partial_{k}G^{rs} \Big(-\iG_{pn}\iG_{sr}-\frac{5}{2} \iG_{pr}\iG_{sn} -\hg_{rn}\tg_{ps}+\frac{1}{2}\hg_{ps}\tg_{nr} \Big)\tg^{km}\ .\nn
\eea
We left a few $\hg_{mn}$ when the terms could not be simplified further. Indeed, as can be seen in \eqref{eq:convrewr}, $\hg_{mn}$ contains two factors of $\iG$ which we cannot always get rid of,
and in that case using $\hg_{mn}$ is shorter.

For later convenience, we would now like to rewrite slightly this expression for $\hR$ in a more compact way.
 To do so, we first consider the following identities:
\bea
\tg^{rs}\tg^{km} \partial_{m}G^{np} \Big(-2\iG_{sn} \partial_{k}\tg_{pr} +\iG_{ps} \partial_{k}\tg_{nr} \Big) =& 
 - \partial_{m}G^{np} \tg^{km}\tg^{rs}\iG_{ps} \partial_{k}\tg_{nr} \label{id1}\\
&\!\!\!\!\!\!\!\!\!\!\!\!\!\!\!\!\!\!\!\!\!\!\!\!\!\!\!\!\!\!\!\! + 2 \partial_{m}G^{np} \tg^{rs}\tg^{km} \Big(-\iG_{sn} \partial_{k}\tg_{pr} + \iG_{ps} \partial_{k}\tg_{nr} \Big) \nn\\
\tg^{qs}\tg^{mr} \partial_{m}G^{np} \Big(2\iG_{sn}\partial_{p}\tg_{qr} -\iG_{ps} \partial_{n}\tg_{qr} \Big) =& 
 \partial_{m}G^{np} \tg^{mr}\tg^{qs}\iG_{ps} \partial_{n}\tg_{qr} \label{id2}\\
&\!\!\!\!\!\!\!\!\!\!\!\!\!\!\!\!\!\!\!\!\!\!\!\!\!\!\!\!\!\!\!\! + 2 \partial_{m}G^{np} \tg^{qs}\tg^{mr} \Big(\iG_{sn}\partial_{p}\tg_{qr} - \iG_{ps} \partial_{n}\tg_{qr} \Big)  \nn\\
\tg^{km} \partial_{m}G^{np}  \partial_{k}G^{rs} \Big(-\hg_{rn}\tg_{ps}+\frac{1}{2}\hg_{ps}\tg_{nr} \Big) =& 
- \frac{1}{2}\tg^{km}\hg_{ps}\tg_{nr} \partial_{m}G^{np}  \partial_{k}G^{rs} \label{id3}\\
& + 4 \tg^{km}\tg_{ps}\hg_{nr} \partial_{m} \tg^{pn}  \partial_{k} \tb^{sr} \nn\\
\partial_{m}G^{np} \Big(\hg_{qn} \partial_{p}G^{qm}  -\frac{1}{2} \hg_{qp} \partial_{n}G^{mq} \Big) =& 
 \frac{1}{2} \hg_{pq} \partial_{m}G^{kp} \partial_{k}G^{mq} \label{id4}\\
& - 4 \hg_{qn} \partial_{m} \tg^{kn} \partial_{k} \tb^{mq} \ ,\nn
\eea
where \eqref{id3} is obtained by comparing $\partial_{m}G^{np}  \partial_{k}G^{rs}$ with $\partial_{m}G^{pn}  \partial_{k}G^{sr}$, and similarly for
\eqref{id4}. The first lines of \eqref{id1}, \eqref{id2}, \eqref{id3}, and \eqref{id4} give on the left-hand side terms of \eqref{R-Ras0}, and on the
right-hand side the terms we want to keep. Therefore, let us show that the sum of the second lines of \eqref{id1}, \eqref{id2}, \eqref{id3}, and \eqref{id4}
 vanishes. More precisely, we can show that
\bea
& \partial_{m}G^{np} \tg^{rs}\tg^{km} \Big(-\iG_{sn} \partial_{k}\tg_{pr} + \iG_{ps} \partial_{k}\tg_{nr} \Big) + 2 \tg^{km}\tg_{ps}\hg_{nr} \partial_{m} \tg^{pn}  \partial_{k} \tb^{sr} = 0 \label{id5}\\
& \partial_{m}G^{np} \tg^{qs}\tg^{mr} \Big(\iG_{sn}\partial_{p}\tg_{qr} - \iG_{ps} \partial_{n}\tg_{qr} \Big) -2 \hg_{qn} \partial_{m} \tg^{kn} \partial_{k} \tb^{mq} = 0 \label{id6} \ . 
\eea
To do so, one can first notice that in the first terms of \eqref{id5} and \eqref{id6}, the symmetric part of $G^{np}$ given by $\tg^{np}$ does not contribute:
the terms vanish thanks to symmetry arguments. Therefore, we are only left with $\partial_{m} \tb^{np}$ in these first terms. Then we can use the following
 symmetric part (see \eqref{ghat}):
\beq
\iG_{(ps)} = \frac{1}{2} \left(\iG_{ps} + \iG_{sp}\right) = \hg_{ps} \ , \label{eq:symG}
\eeq
and this shows that \eqref{id5} and \eqref{id6} do vanish. We conclude that only the right-hand side of the first lines of \eqref{id1}, \eqref{id2},
 \eqref{id3}, and \eqref{id4} remain. We can then rewrite \eqref{R-Ras0} as
\bea
\hR-\tR=& - \partial_{k}\tg_{su} \partial_{m}\tg_{pq} \left(2\tg^{km}\tg^{uq}\tg^{ps}+2\tg^{pq}\tg^{ks}\tg^{mu}+\frac{1}{2}\tg^{uq}\tg^{sm}\tg^{kp}\right) \label{eq:apR-Ras}\\
& -\tg_{pq} \partial_{k}\tb^{pk} \partial_{m}\tb^{qm} -\frac{1}{2}\tg_{pq} \partial_{k}\tb^{qm} \partial_{m}\tb^{pk}  \nn\\
& +2\tg^{km}\tg^{pq} \partial_{k}\partial_{m}\tg_{pq} +2\tg^{km}\iG_{pq} \partial_{k}\partial_{m}G^{qp} \nn\\
& + \partial_{m}G^{vl} \Big( - 2 \tg^{mr} \tg^{ks} \iG_{lv} \del_k \tg_{rs} -\tg^{rs}\tg^{km} \iG_{lv} \partial_{k}\tg_{rs} \nn\\
& \quad +\tg^{ms}\tg^{ru}\iG_{lu} \partial_{v}\tg_{rs} -\tg^{km}\tg^{rs}\iG_{ls} \partial_{k}\tg_{vr} \Big) \nn\\
& + \partial_{m}G^{vl} \Big(\iG_{lq} \partial_{v}G^{qm} +\frac{1}{2} \hg_{lq} \partial_{v} G^{mq} \Big) \nn\\
& - \partial_{m}G^{vl}  \partial_{k}G^{ps}  \frac{1}{2} \tg^{km} \Big(2 \iG_{lv}\iG_{sp} + 5 \iG_{sv}\iG_{lp}  + \hg_{sl}\tg_{pv} \Big)\ .\nn
\eea
It is the same as \eqref{eq:R-Ras}. Let us note that the last formula is also the one we obtain, after computing the expression $\hR-\tR$ without using the
 assumption, and then plugging it in. Finally, one can check that $\hR-\tR$ vanishes for $\tb=0$.

\subsection{The dilaton terms}

We recall that the dilaton $\tp$ is defined as \eqref{tp}
\beq
\tp = \hp + \tfrac{1}{4}{\rm tr}(\ln(\id_d-\tb \tg \tb \tg)) \ .\nn
\eeq
Using for an invertible matrix $A$
\beq
\ln (\det (A))={\rm tr} (\ln(A))\ , \ \partial_{m}\ln (\det (A))={\rm tr} (A^{-1}\partial_m A)\ , \label{eq:lndet}
\eeq
we compute
\bea
\del_m {\rm tr} (\ln (\id_d-\tb\tg \tb\tg ) ) &= {\rm tr} ((\id_d-\tb\tg \tb\tg )^{-1} \partial_m (\id_d-\tb\tg \tb\tg )) \\
&={\rm tr} (G^{-1}\partial_m\tb + \tg^{-1}G^{-1}\tb\partial_m\tg )- {\rm tr} (G^{-T}\partial_m\tb + \tg^{-1}G^{-T}\tb \partial_m\tg)\ .\nn
\eea
Note that \eqref{eq:lndet} is valid as long as $A$ is invertible, whatever signature it has (one may use a complex $\ln$ though if needed). 
In our case, the matrix is clearly invertible since $\id_d-\tb \tg \tb \tg = \hg^{-1} \tg$. In addition, according to footnote \ref{foot:detg}, its determinant is positive.

Using the invariance of the trace under transposition, the cyclicity of the trace and the definition of $G$, one can show that
\beq
{\rm tr} (\tg^{-1}G^{-1}\tb\partial_m\tg )=-{\rm tr} (\tg^{-1}G^{-T}\tb\partial_m\tg )=\iG_{kl}\tg^{ln}\partial_m\tg_{np} \tb^{pk} \ .
\eeq
We introduce for convenience
\beq
A_m=\iG_{kl}\partial_m\tb^{lk} + \iG_{kl}\tg^{ln}\partial_m\tg_{np} \tb^{pk}\ , 
\eeq
and putting everything together, we get
\bea
\d \hp &= \d \tp -\tfrac{1}{2} A_m \d x^m \ ,\\
4 (|\d \hp|^2- |\d \tp|^2) &= 4 (\hg^{km}-\tg^{km}) \del_k \tp \del_m \tp  + \hg^{km} A_k A_m - 4 \hg^{km} A_k \del_m \tp  \ ,\label{dildiff}
\eea
where we mean $|\d \hp|^2=\hg^{km} \del_k \hp \del_m \hp$ and $|\d \tp|^2=\tg^{km} \del_k \tp \del_m \tp$ . It turns out that
\beq
A_m=\tg^{pq} \del_m \tg_{pq}+ \iG_{lk} \del_m G^{kl}\ .
\eeq
Using this and the assumption \eqref{eq:assume}, together with the simplifications \eqref{eq:simplif}, one finally gets
\bea
|\d\hp|^2-|d\tp|^2 &= \tfrac{1}{4} \tg^{km} \tg^{pq} \tg^{uv} \partial_m \tg_{pq} \partial_k \tg_{uv} \vs\\
&\quad + \tfrac{1}{2} \tg^{km} \tg^{pq} \iG_{uv} \partial_m \tg_{pq} \partial_k G^{vu} \vs\nn\\
&\quad + \tfrac{1}{4} \tg^{km} \iG_{pl} \iG_{uv} \partial_m G^{lp} \partial_k G^{vu} \nn\\
&\quad - \tg^{km} \tg^{pq} \partial_k \tg_{pq} \partial_m \tp \nn\\
&\quad - \tg^{km} \iG_{pq} \partial_{k} G^{qp} \partial_m \tp \ , \nn
\eea
which is identical to \eqref{eq:dhpminusdtp}. Note that the assumption makes the term $\del_k \tp \del_m \tp$ in \eqref{dildiff} vanish. We could not get rid
of it otherwise (meaning by another term in the Lagrangian).

\subsection{The \texorpdfstring{$H$}{H}-flux term}

The NSNS $H$-flux is given by the three-form $\hH=\d \hB$. Using \eqref{coefwedge}, and \eqref{eq:convrewr} for the coefficient of $\hB$,  we get
\bea
\label{eq:hhat}
\frac{1}{3} \hH_{kmn} &= \partial_{[k} \hB_{mn]}\\
&= -(G_{-\epsilon}^{-1})_{p [m } \partial_{k} \tb^{p q} (G_{\epsilon}^{-1})_{n] q} - 
2 (G_{-\epsilon}^{-1})_{p [m} \partial_{k} G_{\epsilon}^{p q} \hB_{ n ] q} \ .
\eea
We now want to compute $|\hH|^2$ in which indices are raised with $\hg$ (see \eqref{eq:A2}). Using the above,
 we get
\bea
|\hH|^2=\frac{3}{2} \hg^{k_1 k_2} \hg^{m_1 m_2} \hg^{n_1 n_2} &
\left( (G_{\epsilon_1}^{-1})_{m_1 p_1} \partial_{k_1} \tb^{p_1 q_1} (G_{\epsilon_1}^{-1})_{n_1 q_1} + 
2 (G_{\epsilon_1}^{-1})_{m_1 p_1} \partial_{k_1} G_{\epsilon_1}^{p_1 q_1} \hB_{n_1 q_1} \right) \label{eq:H2}\\
\times &\left( (G_{-\epsilon_2}^{-1})_{p_2 [m_2 } \partial_{k_2} \tb^{p_2 q_2} (G_{\epsilon_2}^{-1})_{n_2] q_2} + 
2 (G_{-\epsilon_2}^{-1})_{p_2 [m_2} \partial_{k_2} G_{\epsilon_2}^{p_2 q_2} \hB_{ n_2 ] q_2} \right) \ ,\nn
\eea
where we antisymmetrize over $k_2,m_2,n_2$ in the second bracket.  Since we contract indices, the antisymmetrization in the first bracket can be neglected. We then get three terms to compute:
\beq
\frac{2}{3}|\hH|^2= (I) + (II) + (III) \ , \label{pH2}
\eeq
where
\bea
(I)&=(\hg^{-1})^3 (G_{\epsilon_1}^{-1})_{m_1 p_1} \partial_{k_1} \tb^{p_1 q_1} (G_{\epsilon_1}^{-1})_{n_1 q_1} (G_{-\epsilon_2}^{-1})_{p_2 [m_2 } \partial_{k_2} \tb^{p_2 q_2} (G_{\epsilon_2}^{-1})_{n_2] q_2}\\
(II)&=4(\hg^{-1})^3 (G_{\epsilon_1}^{-1})_{m_1 p_1} \partial_{k_1} G_{\epsilon_1}^{p_1 q_1} \hB_{n_1 q_1} (G_{-\epsilon_2}^{-1})_{p_2 [m_2 } \partial_{k_2} \tb^{p_2 q_2} (G_{\epsilon_2}^{-1})_{n_2] q_2}\\
(III)&= 4(\hg^{-1})^3(G_{\epsilon_1}^{-1})_{m_1 p_1} \partial_{k_1} G_{\epsilon_1}^{p_1 q_1} \hB_{n_1 q_1} (G_{-\epsilon_2}^{-1})_{p_2 [m_2} \partial_{k_2} G_{\epsilon_2}^{p_2 q_2} \hB_{ n_2 ] q_2}~.
\eea
In this calculation, it is useful to leave $\epsilon=\pm1$ unspecified, and use $G_{\epsilon_1}^{-1}G_{\epsilon_2} \buildrel {\epsilon_1=\epsilon_2} \over \rightarrow 1 $ to contract as many $G_{\epsilon}$ as possible. In order to simplify expressions, let us introduce $D_{\epsilon}^p =G_{\epsilon}^{p q} \partial_q $ , which allows us to write the final result as
\bea
3 (I)=&\left(
\tilde{g}_{p_1 p_2} \tilde{g}_{q_1 q_2} \tilde{g}_{s_1 s_2}-
\tilde{g}_{p_1 s_2} \tilde{g}_{q_1 q_2} \tilde{g}_{s_1 p_2}-
\tilde{g}_{p_1 p_2} \tilde{g}_{q_1 s_2} \tilde{g}_{s_1 q_2}
\right)
D_{\epsilon}^{s_1} \tb^{p_1 q_1} D_{\epsilon}^{s_2} \tb^{p_2 q_2} \label{finalI}\\
3 (II)=&4\left(
\tilde{g}_{p_1 p_2} \tilde{g}_{t_1 q_2} \tilde{g}_{s_1 s_2}-
\tilde{g}_{p_1 s_2} \tilde{g}_{t_1 q_2} \tilde{g}_{s_1 p_2}-
\tilde{g}_{p_1 p_2} \tilde{g}_{t_1 s_2} \tilde{g}_{s_1 q_2}
\right)
\tb^{t_1 t_2} (G_{\epsilon}^{-1})_{q_1 t_2} 
D_{\epsilon}^{s_1} G_{\epsilon}^{p_1 q_1} D_{\epsilon}^{s_2} \tb^{p_2 q_2} \label{finalII}\\
3 (III)=&2 (
\tilde{g}_{p_1 p_2} \tilde{g}_{t_1 t_2} \tilde{g}_{s_1 s_2}-
\tilde{g}_{p_1 s_2} \tilde{g}_{t_1 t_2} \tilde{g}_{s_1 p_2}-
\tilde{g}_{p_1 p_2} \tilde{g}_{t_2 s_2} \tilde{g}_{s_1 t_1} \label{finalIII}\\
&-\tilde{g}_{p_1 t_1} \tilde{g}_{p_2 t_2} \tilde{g}_{s_1 s_2}+
\tilde{g}_{p_1 s_2} \tilde{g}_{t_2 p_2} \tilde{g}_{s_1 t_1}+
\tilde{g}_{p_1 t_1} \tilde{g}_{t_2 s_2} \tilde{g}_{s_1 p_2}) \nn\\
&\left(
\delta_{q_1}^{t_2} - (G_{\epsilon}^{-1})_{q_1 u_2} \tilde{g}^{u_2 t_2}
\right)\left(
\delta_{q_2}^{t_1} - (G_{\epsilon}^{-1})_{q_2 u_1} \tilde{g}^{u_1 t_1}\right) 
D_{\epsilon}^{s_1} G_{\epsilon}^{p_1 q_1} D_{\epsilon}^{s_2} G_{\epsilon}^{p_2 q_2} \ .\nn 
\eea

Using the assumption \eqref{eq:assume} and the simplifications \eqref{eq:simplif}, the previous three terms become
\bea
3(I) =& \tg_{p_1 p_2} \tg_{q_1 q_2} \tg^{s_1 s_2} \del_{s_1} \tb^{p_1 q_1}  \del_{s_2} \tb^{p_2 q_2} 
 - 2 \tg_{q_1 q_2}  \del_{p_2} \tb^{p_1 q_1}  \del_{p_1} \tb^{p_2 q_2} \\
3(II) =& 4 \epsilon \left( \tg_{q_1 q_2} -  (G^{-1}_{\epsilon})_{q_1 q_2} \right) 
\left(-\tg_{p_1 p_2} \tg^{s_1 s_2} \del_{s_2} G_{\epsilon}^{p_1 q_1} \del_{s_1} \tb^{p_2 q_2}  
+ \del_{p_2} G_{\epsilon}^{p_1 q_1} \del_{p_1} \tb^{p_2 q_2}  \right) \\
3(III) =& 2 \del_{p_2} G_{\epsilon}^{p_1 q_1} \del_{p_1} G_{\epsilon}^{p_2 q_2} (\hg_{q_1 q_2} - \tg_{q_1 q_2}) 
+ 2 \tg^{s_1 s_2} \del_{s_1} G_{\epsilon}^{p_1 q_1} \del_{s_2} G_{\epsilon}^{p_2 q_2} \Big( \tg_{p_1 p_2} \tg_{q_1 q_2}  \\
& - \tg_{p_1 q_2} \tg_{q_1 p_2} + 2 \tg_{p_1 q_2} {(G^{-1}_{\epsilon})}_{q_1 p_2} - \tg_{p_1 p_2} \hg_{q_1 q_2} - {(G^{-1}_{\epsilon})}_{q_2 p_1} {(G^{-1}_{\epsilon})}_{q_1 p_2} \Big)  \ , \nn
\eea
where in the last quantity we have used the symmetric part \eqref{eq:symG}. Using in addition \eqref{as1bis} gives the simplified expressions
\bea
3(I) =& \tg_{p_1 p_2} \tg_{q_1 q_2} \tg^{s_1 s_2} \del_{s_1} \tb^{p_1 q_1}  \del_{s_2} \tb^{p_2 q_2} \\
3(II) =& 4 \tg_{p_1 p_2} \tg^{s_1 s_2} \del_{s_1} \tb^{p_2 q_2} \left(-\tg_{q_1 q_2}  \del_{s_2} \tb^{p_1 q_1} + \epsilon {(G^{-1}_{\epsilon})}_{q_1 q_2} \del_{s_2} G_{\epsilon}^{p_1 q_1} \right) \\
3(III) =& 2 (\hg_{q_1 q_2} - \tg_{q_1 q_2}) \del_{p_2} \tg^{p_1 q_1} \del_{p_1} \tg^{p_2 q_2}  
+ 2 \tg^{s_1 s_2} \del_{s_1} G_{\epsilon}^{p_1 q_1} \del_{s_2} G_{\epsilon}^{p_2 q_2} \Big( \tg_{p_1 p_2} \tg_{q_1 q_2} \\
& - \tg_{p_1 q_2} \tg_{q_1 p_2} + 2 \tg_{p_1 q_2} {(G^{-1}_{\epsilon})}_{q_1 p_2} - \tg_{p_1 p_2} \hg_{q_1 q_2} - {(G^{-1}_{\epsilon})}_{q_2 p_1} {(G^{-1}_{\epsilon})}_{q_1 p_2} \Big) \nn\ .
\eea
Note that there was no need to assume \eqref{as1ter} to arrive at the simplified expressions.

We now proceed to some rewriting. First, using the definition of $G_{\epsilon}$, we rewrite
\bea
3(II) =& -4 \tg_{p_1 p_2} \tg_{q_1 q_2} \tg^{s_1 s_2} \del_{s_1} \tb^{p_1 q_1} \del_{s_2} \tb^{p_2 q_2} \\
& + 4 \tg_{p_1 p_2} \tg^{s_1 s_2} \del_{s_1} G_{\epsilon}^{p_1 q_1} \left(\hg_{q_1 q_2} \del_{s_2} G_{\epsilon}^{p_2 q_2} - {(G^{-1}_{\epsilon})}_{q_1 q_2} \del_{s_2} \tg^{p_2 q_2} \right) \ . \nn
\eea
Second, we turn to term $(III)$. Note that
\bea
\tg^{s_1 s_2} \tg_{p_1 q_2} {(G^{-1}_{\epsilon})}_{q_1 p_2} \del_{s_1} G_{\epsilon}^{p_1 q_1} \del_{s_2} G_{\epsilon}^{p_2 q_2} =& \tg^{s_1 s_2} \tg_{p_1 p_2} {(G^{-1}_{\epsilon})}_{q_1 q_2} \del_{s_1} G_{\epsilon}^{p_1 q_1} \del_{s_2} G_{\epsilon}^{q_2 p_2}\\
=& \tg^{s_1 s_2}  \tg_{p_1 p_2} {(G^{-1}_{\epsilon})}_{q_1 q_2} \del_{s_1} G_{\epsilon}^{p_1 q_1} \left(2\del_{s_2} \tg^{p_2 q_2} - \del_{s_2} G_{\epsilon}^{p_2 q_2 } \right) \nn\\
=& \tg^{s_1 s_2}  \tg_{p_1 p_2} \del_{s_1} G_{\epsilon}^{p_1 q_1} \left(2 {(G^{-1}_{\epsilon})}_{q_1 q_2} \del_{s_2} \tg^{p_2 q_2} - \hg_{q_1 q_2} \del_{s_2} G_{\epsilon}^{p_2 q_2 } \right) \ .\nn
\eea
In addition, we note that $\tg_{p_1 p_2} \tg_{q_1 q_2} - \tg_{p_1 q_2} \tg_{q_1 p_2}$ is antisymmetric in $p_1 , q_1$ and in $p_2 , q_2$, so we get
\bea
3(III) =& 4 \tg_{p_1 p_2} \tg_{q_1 q_2} \tg^{s_1 s_2} \del_{s_1} \tb^{p_1 q_1} \del_{s_2} \tb^{p_2 q_2} + 4 \tg^{s_1 s_2}  \tg_{p_1 p_2} \del_{s_1} G_{\epsilon}^{p_1 q_1} \left( {(G^{-1}_{\epsilon})}_{q_1 q_2} \del_{s_2} \tg^{p_2 q_2} - \hg_{q_1 q_2} \del_{s_2} G_{\epsilon}^{p_2 q_2 } \right) \nn\\
& + 2 \tg^{s_1 s_2} \del_{s_1} G_{\epsilon}^{p_1 q_1} \Bigg(2 \tg_{p_1 p_2}  {(G^{-1}_{\epsilon})}_{q_1 q_2} \del_{s_2} \tg^{p_2 q_2} - \left( \tg_{p_1 p_2} \hg_{q_1 q_2} + {(G^{-1}_{\epsilon})}_{q_2 p_1} {(G^{-1}_{\epsilon})}_{q_1 p_2} \right)  \del_{s_2} G_{\epsilon}^{p_2 q_2}\Bigg) \nn\\
& + 2 (\hg_{q_1 q_2} - \tg_{q_1 q_2}) \del_{p_2} \tg^{p_1 q_1} \del_{p_1} \tg^{p_2 q_2} \ , \nn
\eea
where the first line is simply $-3(II)$. Combining all these results, we obtain from (\ref{pH2})
\bea
|\hH|^2 =& \frac{1}{2} \tg_{p_1 p_2} \tg_{q_1 q_2} \tg^{s_1 s_2} \del_{s_1} \tb^{p_1 q_1}  \del_{s_2} \tb^{p_2 q_2} \label{pH2as}\\
& + (\hg_{q_1 q_2} - \tg_{q_1 q_2}) \del_{p_2} \tg^{p_1 q_1} \del_{p_1} \tg^{p_2 q_2} \nn\\
& +  \tg^{s_1 s_2} \del_{s_1} G_{\epsilon}^{p_1 q_1} \Bigg(2\tg_{p_1 p_2}  {(G^{-1}_{\epsilon})}_{q_1 q_2} \del_{s_2} \tg^{p_2 q_2} - \left( \tg_{p_1 p_2} \hg_{q_1 q_2} + {(G^{-1}_{\epsilon})}_{q_2 p_1} {(G^{-1}_{\epsilon})}_{q_1 p_2} \right)  \del_{s_2} G_{\epsilon}^{p_2 q_2}\Bigg) \ .\nn
\eea
We give again this expression in \eqref{eq:Hsquared}, and choose there $\epsilon=+1$.

\end{appendix}
\newpage
\bibliographystyle{JHEP}

\begin{thebibliography}{10}

\bibitem{hmw02}
S.~Hellerman, J.~McGreevy, and B.~Williams, {\it {Geometric constructions of
  nongeometric string theories}},  {\em JHEP} {\bf 01} (2004) 024,
  [\href{http://xxx.lanl.gov/abs/hep-th/0208174}{{\tt hep-th/0208174}}].

\bibitem{dh02}
A.~Dabholkar and C.~Hull, {\it {Duality twists, orbifolds, and fluxes}},  {\em
  JHEP} {\bf 09} (2003) 054,
  [\href{http://xxx.lanl.gov/abs/hep-th/0210209}{{\tt hep-th/0210209}}].

\bibitem{stw05}
J.~Shelton, W.~Taylor, and B.~Wecht, {\it {Nongeometric flux
  compactifications}},  {\em JHEP} {\bf 10} (2005) 085,
  [\href{http://xxx.lanl.gov/abs/hep-th/0508133}{{\tt hep-th/0508133}}].

\bibitem{stw06}
J.~Shelton, W.~Taylor, and B.~Wecht, {\it {Generalized Flux Vacua}},  {\em
  JHEP} {\bf 02} (2007) 095,
  [\href{http://xxx.lanl.gov/abs/hep-th/0607015}{{\tt hep-th/0607015}}].

\bibitem{mpt07}
A.~Micu, E.~Palti, and G.~Tasinato, {\it {Towards Minkowski Vacua in Type II
  String Compactifications}},  {\em JHEP} {\bf 03} (2007) 104,
  [\href{http://xxx.lanl.gov/abs/hep-th/0701173}{{\tt hep-th/0701173}}].

\bibitem{p07}
E.~Palti, {\it {Low Energy Supersymmetry from Non-Geometry}},  {\em JHEP} {\bf
  10} (2007) 011, [\href{http://xxx.lanl.gov/abs/0707.1595}{{\tt
  arXiv:0707.1595}}].

\bibitem{cgm09a}
B.~de~Carlos, A.~Guarino, and J.~M. Moreno, {\it {Flux moduli stabilisation,
  Supergravity algebras and no-go theorems}},  {\em JHEP} {\bf 01} (2010)
  012, [\href{http://xxx.lanl.gov/abs/0907.5580}{{\tt arXiv:0907.5580}}].

\bibitem{cgm09b}
B.~de~Carlos, A.~Guarino, and J.~M. Moreno, {\it {Complete classification of
  Minkowski vacua in generalised flux models}},  {\em JHEP} {\bf 02} (2010)
  076, [\href{http://xxx.lanl.gov/abs/0911.2876}{{\tt arXiv:0911.2876}}].

\bibitem{Blumenhagen:2010hj}
R.~Blumenhagen and E.~Plauschinn, {\it {Nonassociative Gravity in String
  Theory?}},  {\em J. Phys.} {\bf A 44} (2011) 015401,
  [\href{http://xxx.lanl.gov/abs/1010.1263}{{\tt arXiv:1010.1263}}].

\bibitem{Lust:2010iy}
D.~L\"ust, {\it {T-duality and closed string non-commutative (doubled)
  geometry}},  {\em JHEP} {\bf 12} (2010) 084,
  [\href{http://xxx.lanl.gov/abs/1010.1361}{{\tt arXiv:1010.1361}}].

\bibitem{Blumenhagen:2011ph}
R.~Blumenhagen, A.~Deser, D.~L\"ust, E.~Plauschinn, and F.~Rennecke, {\it
  {Non-geometric Fluxes, Asymmetric Strings and Nonassociative Geometry}},
  {\em J. Phys.} {\bf A 44} (2011) 385401, 
  \href{http://xxx.lanl.gov/abs/1106.0316}{{\tt arXiv:1106.0316}}.

\bibitem{gs06}
P.~Grange and S.~Sch\"afer-Nameki, {\it {T-duality with H-flux:
  Non-commutativity, T-folds and G x G structure}},  {\em Nucl. Phys.} {\bf
  B 770} (2007) 123, [\href{http://xxx.lanl.gov/abs/hep-th/0609084}{{\tt
  hep-th/0609084}}].

\bibitem{dh05}
A.~Dabholkar and C.~Hull, {\it {Generalised T-duality and non-geometric
  backgrounds}},  {\em JHEP} {\bf 05} (2006) 009,
  [\href{http://xxx.lanl.gov/abs/hep-th/0512005}{{\tt hep-th/0512005}}].

\bibitem{fww04}
A.~Flournoy, B.~Wecht, and B.~Williams, {\it {Constructing nongeometric vacua
  in string theory}},  {\em Nucl. Phys.} {\bf B 706} (2005) 127,
  [\href{http://xxx.lanl.gov/abs/hep-th/0404217}{{\tt hep-th/0404217}}].

\bibitem{kstt02}
S.~Kachru, M.~B. Schulz, P.~K. Tripathy, and S.~P. Trivedi, {\it {New
  supersymmetric string compactifications}},  {\em JHEP} {\bf 03} (2003) 061,
  [\href{http://xxx.lanl.gov/abs/hep-th/0211182}{{\tt hep-th/0211182}}].

\bibitem{rs08}
R.~A. Reid-Edwards and B.~Spanjaard, {\it {N=4 Gauged Supergravity from
  Duality-Twist Compactifications of String Theory}},  {\em JHEP} {\bf 12}
  (2008) 052, [\href{http://xxx.lanl.gov/abs/0810.4699}{{\tt
  arXiv:0810.4699}}].

\bibitem{mms10}
J.~McOrist, D.~R. Morrison, and S.~Sethi, {\it {Geometries, Non-Geometries, and
  Fluxes}},  \href{http://xxx.lanl.gov/abs/1004.5447}{{\tt arXiv:1004.5447}}.

\bibitem{a11}
D.~Andriot, {\it {Heterotic string from a higher dimensional perspective}},
  \href{http://xxx.lanl.gov/abs/1102.1434}{{\tt arXiv:1102.1434}}.

\bibitem{ms92}
J.~Maharana and J.~H. Schwarz, {\it {Noncompact symmetries in string theory}},
  {\em Nucl. Phys.} {\bf B 390} (1993) 3,
  [\href{http://xxx.lanl.gov/abs/hep-th/9207016}{{\tt hep-th/9207016}}].

\bibitem{hhz10}
O.~Hohm, C.~Hull, and B.~Zwiebach, {\it {Generalized metric formulation of
  double field theory}},  {\em JHEP} {\bf 08} (2010) 008,
  [\href{http://xxx.lanl.gov/abs/1006.4823}{{\tt arXiv:1006.4823}}].

\bibitem{h02}
N.~Hitchin, {\it {Generalized Calabi-Yau manifolds}},  {\em Quart. J. Math. Oxford
  Ser.} {\bf 54} (2003) 281,
  [\href{http://xxx.lanl.gov/abs/math/0209099}{{\tt math/0209099}}].

\bibitem{g04}
M.~Gualtieri, {\it {Generalized complex geometry}},
  \href{http://xxx.lanl.gov/abs/math/0401221}{{\tt math/0401221}}. Ph.D. Thesis
  (Advisor: Nigel Hitchin).

\bibitem{gs07}
P.~Grange and S.~Sch\"afer-Nameki, {\it {Towards mirror symmetry \`a la SYZ for
  generalized Calabi-Yau manifolds}},  {\em JHEP} {\bf 10} (2007) 052,
  [\href{http://xxx.lanl.gov/abs/0708.2392}{{\tt arXiv:0708.2392}}].

\bibitem{gmpw08}
M.~Gra\~na, R.~Minasian, M.~Petrini, and D.~Waldram, {\it {T-duality, Generalized
  Geometry and Non-Geometric Backgrounds}},  {\em JHEP} {\bf 04} (2009) 075,
  [\href{http://xxx.lanl.gov/abs/0807.4527}{{\tt arXiv:0807.4527}}].

\bibitem{a11b}
D.~Andriot, {\it Work in progress}.

\bibitem{lnr03}
D.~A. Lowe, H.~Nastase, and S.~Ramgoolam, {\it {Massive IIA string theory and
  matrix theory compactification}},  {\em Nucl. Phys.} {\bf B 667} (2003) 55,
  [\href{http://xxx.lanl.gov/abs/hep-th/0303173}{{\tt hep-th/0303173}}].

\bibitem{gpr94}
A.~Giveon, M.~Porrati, and E.~Rabinovici, {\it {Target space duality in string
  theory}},  {\em Phys. Rept.} {\bf 244} (1994) 77,
  [\href{http://xxx.lanl.gov/abs/hep-th/9401139}{{\tt hep-th/9401139}}].

\bibitem{t10}
D.~C. Thompson, {\it {T-duality Invariant Approaches to String Theory}},
  \href{http://xxx.lanl.gov/abs/1012.4393}{{\tt arXiv:1012.4393}}.

\bibitem{syz96}
A.~Strominger, S.-T. Yau, and E.~Zaslow, {\it {Mirror symmetry is T duality}},
  {\em Nucl. Phys.} {\bf B 479} (1996) 243,
  [\href{http://xxx.lanl.gov/abs/hep-th/9606040}{{\tt hep-th/9606040}}].

\bibitem{kv96}
A.~Kumar and C.~Vafa, {\it {U manifolds}},  {\em Phys. Lett.} {\bf B 396} (1997)
  85, [\href{http://xxx.lanl.gov/abs/hep-th/9611007}{{\tt
  hep-th/9611007}}].

\bibitem{hc03}
C.~M. Hull and A.~Catal-Ozer, {\it {Compactifications with S-duality twists}},
  {\em JHEP} {\bf 10} (2003) 034,
  [\href{http://xxx.lanl.gov/abs/hep-th/0308133}{{\tt hep-th/0308133}}].

\bibitem{r06}
R.~A. Reid-Edwards, {\it {Geometric and non-geometric compactifications of IIB
  supergravity}},  {\em JHEP} {\bf 12} (2008) 043,
  [\href{http://xxx.lanl.gov/abs/hep-th/0610263}{{\tt hep-th/0610263}}].

\bibitem{h04}
C.~M. Hull, {\it {A Geometry for non-geometric string backgrounds}},  {\em JHEP}
  {\bf 10} (2005) 065, [\href{http://xxx.lanl.gov/abs/hep-th/0406102}{{\tt
  hep-th/0406102}}].

\bibitem{hm06}
E.~Hackett-Jones and G.~Moutsopoulos, {\it {Quantum mechanics of the doubled
  torus}},  {\em JHEP} {\bf 10} (2006) 062,
  [\href{http://xxx.lanl.gov/abs/hep-th/0605114}{{\tt hep-th/0605114}}].

\bibitem{h06b}
C.~M. Hull, {\it {Doubled Geometry and T-Folds}},  {\em JHEP} {\bf 07} (2007)
  080, [\href{http://xxx.lanl.gov/abs/hep-th/0605149}{{\tt hep-th/0605149}}].

\bibitem{bct07}
D.~S. Berman, N.~B. Copland, and D.~C. Thompson, {\it {Background Field
  Equations for the Duality Symmetric String}},  {\em Nucl. Phys.} {\bf B 791}
  (2008) 175, [\href{http://xxx.lanl.gov/abs/0708.2267}{{\tt
  arXiv:0708.2267}}].

\bibitem{bt07}
D.~S. Berman and D.~C. Thompson, {\it {Duality Symmetric Strings, Dilatons and
  O(d,d) Effective Actions}},  {\em Phys. Lett.} {\bf B 662} (2008) 279,
  [\href{http://xxx.lanl.gov/abs/0712.1121}{{\tt arXiv:0712.1121}}].

\bibitem{c11}
N.~B. Copland, {\it {Connecting T-duality invariant theories}},
  \href{http://xxx.lanl.gov/abs/1106.1888}{{\tt arXiv:1106.1888}}.

\bibitem{glmw02}
S.~Gurrieri, J.~Louis, A.~Micu, and D.~Waldram, {\it {Mirror symmetry in
  generalized Calabi-Yau compactifications}},  {\em Nucl. Phys.} {\bf B 654}
  (2003) 61, [\href{http://xxx.lanl.gov/abs/hep-th/0211102}{{\tt
  hep-th/0211102}}].

\bibitem{fmt03}
S.~Fidanza, R.~Minasian, and A.~Tomasiello, {\it {Mirror symmetric SU(3)
  structure manifolds with NS fluxes}},  {\em Commun. Math. Phys.} {\bf 254}
  (2005) 401, [\href{http://xxx.lanl.gov/abs/hep-th/0311122}{{\tt
  hep-th/0311122}}].

\bibitem{st08}
W.~Schulgin and J.~Troost, {\it {Backreacted T-folds and non-geometric regions
  in configuration space}},  {\em JHEP} {\bf 12} (2008) 098,
  [\href{http://xxx.lanl.gov/abs/0808.1345}{{\tt arXiv:0808.1345}}].

\bibitem{fw05}
A.~Flournoy and B.~Williams, {\it {Nongeometry, duality twists, and the
  worldsheet}},  {\em JHEP} {\bf 01} (2006) 166,
  [\href{http://xxx.lanl.gov/abs/hep-th/0511126}{{\tt hep-th/0511126}}].

\bibitem{hw06}
S.~Hellerman and J.~Walcher, {\it {Worldsheet CFTs for Flat Monodrofolds}},
  \href{http://xxx.lanl.gov/abs/hep-th/0604191}{{\tt hep-th/0604191}}.

\bibitem{mw86}
M.~T. Mueller and E.~Witten, {\it {Twisting Toroidally Compactified Heterotic
  Strings with Enlarged Symmetry Groups}},  {\em Phys. Lett.} {\bf B 182} (1986)
  28.

\bibitem{nsv87}
K.~S. Narain, M.~H. Sarmadi, and C.~Vafa, {\it {Asymmetric Orbifolds}},  {\em
  Nucl. Phys.} {\bf B 288} (1987) 551.

\bibitem{w07}
B.~Wecht, {\it {Lectures on Nongeometric Flux Compactifications}},  {\em
  Class. Quant. Grav.} {\bf 24} (2007) S773,
  [\href{http://xxx.lanl.gov/abs/0708.3984}{{\tt arXiv:0708.3984}}].

\bibitem{b87}
T.~H. Buscher, {\it {A Symmetry of the String Background Field Equations}},  {\em
  Phys. Lett.} {\bf B 194} (1987) 59.

\bibitem{b88}
T.~H. Buscher, {\it {Path Integral Derivation of Quantum Duality in Nonlinear
  Sigma Models}},  {\em Phys. Lett.} {\bf B 201} (1988) 466.

\bibitem{km99}
N.~Kaloper and R.~C. Myers, {\it {The Odd story of massive supergravity}},
  {\em JHEP} {\bf 05} (1999) 010,
  [\href{http://xxx.lanl.gov/abs/hep-th/9901045}{{\tt hep-th/9901045}}].

\bibitem{Derendinger:2004jn}
J.-P. Derendinger, C.~Kounnas, P.~Petropoulos, and F.~Zwirner, {\it
  {Superpotentials in IIA compactifications with general fluxes}},  {\em
  Nucl. Phys.} {\bf B 715} (2005) 211,
  [\href{http://xxx.lanl.gov/abs/hep-th/0411276}{{\tt hep-th/0411276}}].

\bibitem{df05}
G.~Dall'Agata and S.~Ferrara, {\it {Gauged supergravity algebras from twisted
  tori compactifications with fluxes}},  {\em Nucl. Phys.} {\bf B 717} (2005)
  223, [\href{http://xxx.lanl.gov/abs/hep-th/0502066}{{\tt
  hep-th/0502066}}].

\bibitem{hr05}
C.~M. Hull and R.~A. Reid-Edwards, {\it {Flux compactifications of string theory on
  twisted tori}},  {\em Fortsch. Phys.} {\bf 57} (2009) 862,
  [\href{http://xxx.lanl.gov/abs/hep-th/0503114}{{\tt hep-th/0503114}}].

\bibitem{dpst07}
G.~Dall'Agata, N.~Prezas, H.~Samtleben, and M.~Trigiante, {\it {Gauged
  Supergravities from Twisted Doubled Tori and Non-Geometric String
  Backgrounds}},  {\em Nucl. Phys.} {\bf B 799} (2008) 80,
  [\href{http://xxx.lanl.gov/abs/0712.1026}{{\tt arXiv:0712.1026}}].

\bibitem{hr07}
C.~M. Hull and R.~A. Reid-Edwards, {\it {Gauge symmetry, T-duality and doubled
  geometry}},  {\em JHEP} {\bf 08} (2008) 043,
  [\href{http://xxx.lanl.gov/abs/0711.4818}{{\tt arXiv:0711.4818}}].

\bibitem{eg95}
M.~Evans and I.~Giannakis, {\it {T-duality in arbitrary string backgrounds}},
  {\em Nucl. Phys.} {\bf B 472} (1996) 139,
  [\href{http://xxx.lanl.gov/abs/hep-th/9511061}{{\tt hep-th/9511061}}].

\bibitem{h06a}
C.~M. Hull, {\it {Global aspects of T-duality, gauged sigma models and T-folds}},
  {\em JHEP} {\bf 10} (2007) 057,
  [\href{http://xxx.lanl.gov/abs/hep-th/0604178}{{\tt hep-th/0604178}}].

\bibitem{j11}
S.~Jensen, {\it {The KK-Monopole/NS5-Brane in Doubled Geometry}},  {\em JHEP}
  {\bf 07} (2011) 088, [\href{http://xxx.lanl.gov/abs/1106.1174}{{\tt
  arXiv:1106.1174}}].

\bibitem{h09}
N.~Halmagyi, {\it {Non-geometric Backgrounds and the First Order String Sigma
  Model}},  \href{http://xxx.lanl.gov/abs/0906.2891}{{\tt arXiv:0906.2891}}.

\bibitem{agmp10}
D.~Andriot, E.~Goi, R.~Minasian, and M.~Petrini, {\it {Supersymmetry breaking
  branes on solvmanifolds and de Sitter vacua in string theory}},  {\em JHEP}
  {\bf 05} (2011) 028, [\href{http://xxx.lanl.gov/abs/1003.3774}{{\tt
  arXiv:1003.3774}}].

\bibitem{e06}
I.~T. Ellwood, {\it {NS-NS fluxes in Hitchin's generalized geometry}},  {\em
  JHEP} {\bf 12} (2007) 084,
  [\href{http://xxx.lanl.gov/abs/hep-th/0612100}{{\tt hep-th/0612100}}].

\bibitem{mpz06}
R.~Minasian, M.~Petrini, and A.~Zaffaroni, {\it {Gravity duals to deformed SYM
  theories and Generalized Complex Geometry}},  {\em JHEP} {\bf 12} (2006)
  055, [\href{http://xxx.lanl.gov/abs/hep-th/0606257}{{\tt hep-th/0606257}}].

\bibitem{lm05}
O.~Lunin and J.~M. Maldacena, {\it {Deforming field theories with U(1) x U(1)
  global symmetry and their gravity duals}},  {\em JHEP} {\bf 05} (2005) 033,
  [\href{http://xxx.lanl.gov/abs/hep-th/0502086}{{\tt hep-th/0502086}}].

\bibitem{h08}
N.~Halmagyi, {\it {Non-geometric String Backgrounds and Worldsheet Algebras}},
  {\em JHEP} {\bf 07} (2008) 137,
  [\href{http://xxx.lanl.gov/abs/0805.4571}{{\tt arXiv:0805.4571}}].

\bibitem{irw07}
M.~Ihl, D.~Robbins, and T.~Wrase, {\it {Toroidal orientifolds in IIA with
  general NS-NS fluxes}},  {\em JHEP} {\bf 08} (2007) 043,
  [\href{http://xxx.lanl.gov/abs/0705.3410}{{\tt arXiv:0705.3410}}].

\bibitem{glw05}
M.~Gra\~na, J.~Louis, and D.~Waldram, {\it {Hitchin functionals in N=2
  supergravity}},  {\em JHEP} {\bf 01} (2006) 008,
  [\href{http://xxx.lanl.gov/abs/hep-th/0505264}{{\tt hep-th/0505264}}].

\bibitem{bg06}
I.~Benmachiche and T.~W. Grimm, {\it {Generalized N=1 orientifold
  compactifications and the Hitchin functionals}},  {\em Nucl. Phys.} {\bf B 748}
  (2006) 200, [\href{http://xxx.lanl.gov/abs/hep-th/0602241}{{\tt
  hep-th/0602241}}].

\bibitem{glw06}
M.~Gra\~na, J.~Louis, and D.~Waldram, {\it {SU(3) x SU(3) compactification and
  mirror duals of magnetic fluxes}},  {\em JHEP} {\bf 04} (2007) 101,
  [\href{http://xxx.lanl.gov/abs/hep-th/0612237}{{\tt hep-th/0612237}}].

\bibitem{acfi06}
G.~Aldazabal, P.~G. Camara, A.~Font, and L.~Ib\'a\~nez, {\it {More dual fluxes and
  moduli fixing}},  {\em JHEP} {\bf 05} (2006) 070,
  [\href{http://xxx.lanl.gov/abs/hep-th/0602089}{{\tt hep-th/0602089}}].

\bibitem{rw07}
D.~Robbins and T.~Wrase, {\it {D-terms from generalized NS-NS fluxes in type
  II}},  {\em JHEP} {\bf 12} (2007) 058,
  [\href{http://xxx.lanl.gov/abs/0709.2186}{{\tt arXiv:0709.2186}}].

\bibitem{bhm07}
D.~M. Belov, C.~M. Hull, and R.~Minasian, {\it {T-duality, gerbes and loop
  spaces}},  \href{http://xxx.lanl.gov/abs/0710.5151}{{\tt arXiv:0710.5151}}.

\bibitem{bbvw07}
K.~Becker, M.~Becker, C.~Vafa, and J.~Walcher, {\it {Moduli Stabilization in
  Non-Geometric Backgrounds}},  {\em Nucl. Phys.} {\bf B 770} (2007) 1,
  [\href{http://xxx.lanl.gov/abs/hep-th/0611001}{{\tt hep-th/0611001}}].

\bibitem{hktt08}
M.~P. Hertzberg, S.~Kachru, W.~Taylor, and M.~Tegmark, {\it {Inflationary
  Constraints on Type IIA String Theory}},  {\em JHEP} {\bf 12} (2007) 095,
  [\href{http://xxx.lanl.gov/abs/0711.2512}{{\tt arXiv:0711.2512}}].

\bibitem{hz09a}
C.~Hull and B.~Zwiebach, {\it {Double Field Theory}},  {\em JHEP} {\bf 09}
  (2009) 099, [\href{http://xxx.lanl.gov/abs/0904.4664}{{\tt
  arXiv:0904.4664}}].

\bibitem{hz09b}
C.~Hull and B.~Zwiebach, {\it {The Gauge algebra of double field theory and
  Courant brackets}},  {\em JHEP} {\bf 09} (2009) 090,
  [\href{http://xxx.lanl.gov/abs/0908.1792}{{\tt arXiv:0908.1792}}].

\bibitem{hhz10a}
O.~Hohm, C.~Hull, and B.~Zwiebach, {\it {Background independent action for
  double field theory}},  {\em JHEP} {\bf 07} (2010) 016,
  [\href{http://xxx.lanl.gov/abs/1003.5027}{{\tt arXiv:1003.5027}}].

\bibitem{ms07}
F.~Marchesano and W.~Schulgin, {\it {Non-geometric fluxes as supergravity
  backgrounds}},  {\em Phys. Rev.} {\bf D 76} (2007) 041901,
  [\href{http://xxx.lanl.gov/abs/0704.3272}{{\tt arXiv:0704.3272}}].

\bibitem{acr09}
G.~Aldazabal, P.~G. Camara, and J.~Rosabal, {\it {Flux algebra, Bianchi
  identities and Freed-Witten anomalies in F-theory compactifications}},  {\em
  Nucl. Phys.} {\bf B 814} (2009) 21,
  [\href{http://xxx.lanl.gov/abs/0811.2900}{{\tt arXiv:0811.2900}}].

\bibitem{dlr10}
G.~Dibitetto, R.~Linares, and D.~Roest, {\it {Flux Compactifications, Gauge
  Algebras and De Sitter}},  {\em Phys. Lett.} {\bf B 688} (2010) 96,
  [\href{http://xxx.lanl.gov/abs/1001.3982}{{\tt arXiv:1001.3982}}].

\bibitem{h07}
C.~M. Hull, {\it {Generalised Geometry for M-Theory}},  {\em JHEP} {\bf 07}
  (2007) 079, [\href{http://xxx.lanl.gov/abs/hep-th/0701203}{{\tt
  hep-th/0701203}}].

\bibitem{pw08}
P.~P. Pacheco and D.~Waldram, {\it {M-theory, exceptional generalised geometry
  and superpotentials}},  {\em JHEP} {\bf 09} (2008) 123,
  [\href{http://xxx.lanl.gov/abs/0804.1362}{{\tt arXiv:0804.1362}}].

\bibitem{glsw09}
M.~Gra\~na, J.~Louis, A.~Sim, and D.~Waldram, {\it {E7(7) formulation of N=2
  backgrounds}},  {\em JHEP} {\bf 07} (2009) 104,
  [\href{http://xxx.lanl.gov/abs/0904.2333}{{\tt arXiv:0904.2333}}].

\bibitem{aacg10}
G.~Aldazabal, E.~Andres, P.~G. Camara, and M.~Gra\~na, {\it {U-dual fluxes and
  Generalized Geometry}},  {\em JHEP} {\bf 11} (2010) 083,
  [\href{http://xxx.lanl.gov/abs/1007.5509}{{\tt arXiv:1007.5509}}].

\bibitem{go11}
M.~Gra\~na and F.~Orsi, {\it {N=1 vacua in Exceptional Generalized Geometry}},
  \href{http://xxx.lanl.gov/abs/1105.4855}{{\tt arXiv:1105.4855}}.

\bibitem{bp10}
D.~S. Berman and M.~J. Perry, {\it {Generalized Geometry and M theory}},  {\em
  JHEP} {\bf 06} (2011) 074, [\href{http://xxx.lanl.gov/abs/1008.1763}{{\tt
  arXiv:1008.1763}}].

\bibitem{bgp11}
D.~S. Berman, H.~Godazgar, and M.~J. Perry, {\it {SO(5,5) duality in M-theory
  and generalized geometry}},  {\em Phys. Lett.} {\bf B 700} (2011) 65,
  [\href{http://xxx.lanl.gov/abs/1103.5733}{{\tt arXiv:1103.5733}}].

\end{thebibliography}
\providecommand{\href}[2]{#2}\begingroup\raggedright\endgroup

\end{document}